\documentclass[preprint,aps,12pt,showpacs,nofootinbib,tightenlines]{revtex4}
\usepackage{amsmath}
\usepackage{amssymb}
\usepackage{epsfig}
\usepackage{graphicx}
\begin{document}
\newcommand{\beq}{\begin{eqnarray}}
\newcommand{\eeq}{\end{eqnarray}}
\def \epjc{  Eur. Phys. J. C }
\def \jpg{  J. Phys. G }
\def \npb{  Nucl. Phys. B }
\def \plb{  Phys. Lett. B }
\def \prd{  Phys. Rev. D }
\def \prl{  Phys. Rev. Lett.  }
\def \pr{   Phys. Rep. }
\def \rmp{  Rev. Mod. Phys. }
\def \ptp{  Prog. Theor. Phys. }

\newcommand{\non}{\nonumber\\ }
\title{Charmless $B \to PV, VV $ decays and new physics
effects in the mSUGRA model}
\author{Wenjuan Zou}
\email{zouwenjuan@email.njnu.edu.cn}
\affiliation{Department of Physics and Institute of Theoretical Physics, 
Nanjing Normal University, Nanjing, Jiangsu 210097, P.R.China}
\author{ Zhenjun Xiao}
\email{xiaozhenjun@njnu.edu.cn}
\affiliation{Department of Physics and Institute of Theoretical Physics, 
Nanjing Normal University, Nanjing, Jiangsu 210097, P.R.China}
\affiliation{ CCAST(World Laboratory), P.O.Box 8730, Beijing 100080, China}
\date{\today}
\begin{abstract}
By employing the QCD factorization approach, we calculate the new
physics contributions to the branching radios of the two-body
charmless $ B \to PV$ and $B \to VV$ decays in the framework of
the minimal supergravity (mSUGRA) model.
we choose three typical sets of the mSUGRA input parameters in which
the Wilson coefficient $C_{7\gamma}(m_b)$ can be either SM-like ( the case A and C) or
has a flipped-sign (the case B). We found numerically that
(a) the SUSY contributions are always very small for both case A and C;
(b) for those tree-dominated decays, the SUSY contributions in case B
are also very small;
(c) for those QCD penguin-dominated decay modes, the SUSY
contributions in case B can be significant, and can provide an enhancement about
$30\% \sim 260\%$ to the branching ratios of $B \to K^*(\pi,\phi,\rho)$ and $K \phi$
decays, but a reduction about $30\% \sim 80\%$ to
$ B\to K(\rho, \omega)$ decays; 
and (d) the large SUSY contributions in the case B may be masked 
by the large theoretical errors dominated by the uncertainty from
our ignorance of calculating the annihilation contributions 
in the QCD factorization approach. 
\end{abstract}


\maketitle

\section{Introduction}\label{sec:int}

Along with the excellent performance of the B factory experiments
\cite{belle,babar}
and the forthcoming LHCb and other B meson related experiments\cite{lhcb},
huge amount of B data with great precision are expected in
the following five to ten years.
The precision measurements of the B meson system can provide
an insight into very high energy scales via the indirect loop
effects of the new physics beyond the standard model (SM)
\cite{slac504,hurth03}. Although currently available data
agree well with the SM predictions when considering the still large
uncertainties both in theory and experiments, we generally believe
that the B-factories can at least detect the first signals of new
physics if it is there.

As for the charmless hadronic two-body B decays considered here,
people indeed have found some deviations from the SM expectations,
such as the unexpected large branching ratios of $B \to K \eta'$
and $B \to \pi^0 \pi^0$ decay modes, the $"K \pi"$ puzzle
\cite{hurth03,gr03} and the so-called $\phi K_s$ anomaly at
$3.8\sigma$ level \cite{phiks}. Although not convincing, these
potential deviations may be considered as the first hints of new
physics beyond the SM in B meson experiments.

On the theory side, by employing the low energy effective
Hamiltonian and various factorization hypothesis the charmless
B meson decays have been widely studied in
the framework of the SM \cite{agc98,yhb99,dhj02,bbns01,bn03a,bn03b,pqcd}.
The possible new physics contributions to B meson decays induced by
loop diagrams involving various new particles
have been studied extensively, for example, in the
Technicolor models\cite{tc2}, the two-Higgs-doublet
models\cite{etapm3,2hdm1} and the supersymmetric models
\cite{hmdx98,tyy96,tyy97,mssm,khalil}.

Among the various new physics models, the minimal supergravity
(mSUGRA) model \cite{msugra} is a constrained minimal
supersymmetric standard model \cite{as99} which has only five free
parameters ( $ \tan\beta, m_{\frac{1}{2}}, m_{0},A_{0}$, and
$sign(\mu)$ ) at the high energy scale.  In
Refs.~\cite{hmdx98,tyy96,tyy97,huang03}, for example, the authors
studied the rare decays $B\to X_s \gamma$, $B \to X_s l\bar{l}$,
$B \to l^+ l^-$ and the $B^{0}-\bar{B^{0} }$ mixing in the mSUGRA
model, and found some constraints on the parameter space of this
model.

In a recent paper\cite{zw04}, we calculated the SUSY contributions
to the branching ratios of the twenty one $B \to PP$ (P stands for
pseudo-scalar light meson) decay modes in the mSUGRA model by
employing the QCD factorization approach. We deduced analytically
the contributions given by the new penguin diagrams induced by
gluinos, charged-Higgs bosons, charginos and neutralinos, and
obtained the analytical expressions of the SUSY corrections to the
Wilson coefficients. By considering the current constraints, such
as the branching ratio of the inclusive radiative B-decay,
$C_{7\gamma}(m_b)$ can be either SM-like or sign-flipped comparing
with that in the SM. At the considered parameter point where the
SUSY contributions can make the sign of $C_{7\gamma}(m_b)$
reversed with respect to the SM one, we found that (a) the SUSY
corrections to the Wilson coefficients $C_{7\gamma}(M_W)$ and
$C_{8g}(M_W)$ can be rather large; (b) the SUSY enhancements to
those penguin-dominated decays can be as large as $30\%$-$50\%$;
(c)for $B \to K \pi$ decays, the inclusion of the SUSY
enhancements can improve the consistency of the theoretical
prediction with the data; and (d) the large SUSY contributions to
$B \to K \eta'$ decays will help us to give a new physics
interpretation for the so-called $`` K \eta^{'}  "$ puzzle.

In this paper, we will extend the previous work \cite{zw04} to the
cases for thirty nine $B \to P V$ decays and nineteen $B \to VV$
decay modes (here V stands for the light vector meson). This paper
is organized as follows. In Sec.~\ref{sec:msugra} and
Sec.~\ref{sec:qcdf}, we give a brief review for the mSUGRA model
and the QCD factorization approach for two-body B meson decays.
In Sec.~\ref{sec:pv} and
Sec.~\ref{sec:vv}, we present the related formulae and calculate
the CP-averaged branching ratios for $B\to PV$ and $B \to VV$
decays, respectively. The summary and some discussions are
included in the final section.

\section{The mSUGRA model and the new physics contributions}
\label{sec:msugra}

In this section,  we first recapitulate the basic theoretical
framework of the mSUGRA model to set up the notation and then calculate the
SUSY corrections to the Wilson coefficients in the mSUGRA model.

\subsection{Outline of the mSUGRA model}

The most general superpotential compatible with gauge invariance,
renormalizability and R-parity conserving in MSSM can be written as \cite{as99}:
\begin{eqnarray}
{\cal W}=\varepsilon_{\alpha\beta}\left
[f_{Uij}Q_{i}^{\alpha}H_{2}^{\beta}U_{j}
                           +f_{Dij}H_{1}^{\alpha}Q_{i}^{\beta}D_{j}
                           +f_{Eij}H_{1}^{\alpha}L_{i}^{\beta}E_{j}
                           -\mu H_{1}^{\alpha}H_{2}^{\beta} \right ]
\end{eqnarray}
Where $Q,U,D,L,E,H_1$ and $H_2$ are  chiral superfields. $f_{D}$,
$f_{U}$ and $f_{E}$ are Yukawa coupling constants for down-type,
up-type quarks, and leptons, respectively. The suffixes
$\alpha,\beta=1,2$ are SU(2) indices and i,j=1,2,3 are generation
indices. $\varepsilon_{\alpha\beta}$ is the antisymmetric tensor
with $ \varepsilon_{12}=1$.

In addition to the SUSY invariant terms, a set of terms which
explicitly but softly break SUSY should be added to the
supersymmetric Lagrangian. A general form of the soft
SUSY-breaking terms is given as:
\begin{eqnarray}
-{\cal L}_{soft}&=&
    \left (m^{2}_{Q}\right )_{ij}\tilde{q}^{+}_{Li}\tilde{q}_{Lj}
   +\left(m^{2}_U\right )_{ij}\tilde{u}^{*}_{Ri}\tilde{u}_{Rj}
   +\left(m^{2}_D\right )_{ij}\tilde{d}^{*}_{Ri}\tilde{d}_{Rj}
   +\left(m^{2}_L\right )_{ij}\tilde{l}^{+}_{Li}\tilde{l}_{Lj}\non
&\ \ &
   +\left(m^{2}_E\right )_{ij}\tilde{e}^{*}_{Ri}\tilde{e}_{Rj}
   +\Delta^{2}_{1}h_{1}^{+}h_{1}+\Delta^{2}_{2}h_{2}^{+}h_{2} \non
&\ \ &
   +\varepsilon_{\alpha\beta}
   \left [A_{Uij}\tilde{q}^{\alpha}_{Li}h^{\beta}_{2}\tilde{u}^{*}_{Rj}
   +A_{Dij}h^{\alpha}_{1}\tilde{q}^{\beta}_{Li}\tilde{d}^{*}_{Rj}
   +A_{Eij}h^{\alpha}_{1}\tilde{l}^{\beta}_{Li}\tilde{e}^{*}_{Rj}
   +B\mu h^{\alpha}_{1}h^{\beta}_{2}\right ]\non
&\ \ &
   +\frac{1}{2}m_{\tilde{B}}\tilde{B}\tilde{B}
   +\frac{1}{2}m_{\tilde{W}}\tilde{W}\tilde{W}
   +\frac{1}{2}m_{\tilde{G}}\tilde{G}\tilde{G} + H.C.
\label{eq:lsoft}
\end{eqnarray}

where $\tilde{q}_{Li}$, $\tilde{u}^{*}_{Ri}$,
$\tilde{d}^{*}_{Ri}$, $\tilde{l}_{Li}$, $\tilde{e}^{*}_{Ri}$, and
$h_1$ and $h_2$ are scalar components of chiral superfields $Q_i$,
$U_i$, $ D_{i}$, $L_{i}$, $E_{i}$, $H_1$, and $H_2$ respectively,
and $\tilde{B}$, $\tilde{W}$, and $\tilde{G}$ are $ U(1)_Y$,
$SU(2)_L$, and $ SU(3)_C $ gauge fermions. And the terms appeared
in Eq.(\ref{eq:lsoft}) are the mass terms for the scalar fermions,
mass and bilinear terms for the Higgs bosons, trilinear coupling
terms between sfermions and Higgs bosons, and mass terms for the
gluinos, Winos and binos, respectively.

In the mSUGRA model, a set of assumptions are added to the MSSM.
One underlying assumption is that SUSY-breaking occurs in a hidden
sector which communicates with the visible sector only through
gravitational interactions.  The free parameters in the MSSM are
assumed to obey a set of boundary conditions at the Plank or GUT
scale:
\begin{eqnarray}
 \alpha_{1}&=&\alpha_{2}=\alpha_{3}=\alpha_{X}, \non
(m^{2}_{Q})_{ij}&=&
(m^{2}_{U})_{ij}=(m^{2}_{D})_{ij}=(m^{2}_{L})_{ij}
=(m^{2}_{E})_{ij}=(m^{2}_{0})\delta_{ij}, \non
\Delta^{2}_{1}&=&\Delta^{2}_{2}=m^{2}_{0}, \non
A_{Uij}&=&f_{Uij}A_{0},\ \ A_{Dij}=f_{Dij}A_{0}, \ \
A_{Eij}=f_{Eij}A_{0}, \non m_{\tilde{B}}&=&
m_{\tilde{W}}=m_{\tilde{G}}=m_{\frac{1}{2}}
\end{eqnarray}
where $\alpha_{i}=g^2_i/(4\pi)$, while  $g_{i}$ (i=1,2,3) denotes
the coupling constant of the $U(1)_Y$, $SU(2)_L$, $SU(3)_C$ gauge
group, respectively. The unification of them is verified according
to the experimental results from LEP1\cite{pdg04} and  can be
fixed at the Grand Unification Scale $M_{GUT}\sim 2\times
10^{16}Gev$.

Besides the three parameters $ m_{\frac{1}{2}}$, $m_{0}$ and
$A_{0}$, the bilinear coupling B and the supersymmetric Higgs(ino)
mass parameter $\mu$ in the supersymmetric sector should be
determined. By requiring the radiative electroweak
symmetry-breaking (EWSB) takes place at the low energy scale, both
of them are obtained except for the sign of $\mu$.
At this stage, only four continuous free parameters and an unknown sign are
left in the mSUGRA model \cite{kah}:
\begin{eqnarray}
\tan\beta, m_{\frac{1}{2}}, m_{0},A_{0},sign(\mu)
\end{eqnarray}
In present analysis, we assume that they are all real parameters.
Therefore there are  no new CP-violating complex phase introduced
other than that in the Cabbibo-Kabayashi-Maskawa(CKM) quark mixing
matrix \cite{ckm}. Once the five free parameters of the mSUGRA
model are determined, all other parameters at the electroweak
scale are then obtained through the GUT universality condition and
the renormalization group equation (RGE) evolution. Like in
Ref.~\cite{zw04}, we here also use the Fortran code - Suspect
version 2.1 \cite{ajg02}- to calculate the SUSY and Higgs particle
spectrum \footnote{For more details, one can see discussions in
Ref.~\cite{zw04} and references therein.}.

\subsection{New physics effects in the mSUGRA model}

As is well-known, the low energy effective Hamiltonian for the
quark level three-body decay $ b\to qq^{'}\bar{q^{'}}$
$(q\in\{d,s\}, q^{'}\in \{ u,c,d,s,b\})$ at the scale $\mu \sim m_b$ reads
\begin{eqnarray}
{\cal H}_{eff}&=&\frac{G_{F}}{\sqrt{2}}\left\{\sum_{i=1}^{2}C_{i}(\mu)
\left
[V_{ub}V_{uq}^{*}O_{i}^{u}(\mu)+V_{cb}V_{cq}^{*}O_{i}^{c}(\mu)\right
] \right.\non
 &  & \left. -V_{tb}V_{tq}^{*}\sum_{j=3}^{10}C_{j}(\mu)O_{j}(\mu)
             -V_{tb}V_{tq}^{*}\left [C_{7\gamma}(\mu)O_{7\gamma}(\mu)+C_{8g}(\mu)O_{8g}(\mu)
             \right ] \right \}
\label{eq:eff}
\end{eqnarray}
where $V_{pb}V_{pq}^{*}$ with $q=d,s$ is the products of elements
of the CKM matrix \cite{ckm}. The definitions and the explicit
expressions of the operators $O_{i}$ $(i=1\sim10,7\gamma,8g)$ and
the corresponding Wilson coefficients $C_i$ can be found in
Ref.~\cite{gam96}. In the SM, the Wilson coefficients appeared in
eq.(\ref{eq:eff}) are currently known at next-to-leading order
(NLO) and can be found easily in Ref.~\cite{gam96}.

In the mSUGRA model, the new physics effects on the rare
B meson decays will manifest themselves through two channels.
\begin{itemize}
\item
The new physics contributions to the Wilson coefficients of the same
operators involved in the SM calculation;

\item
The other is the new physics contributions to the Wilson coefficients of the
new operators such as the operators with opposite chiralities with $O_i$ appeared
in eq.(\ref{eq:eff}).
\end{itemize}
In the SM, the latter is absent because they are suppressed by the
ratio $m_{s}/m_{b}$. In the mSUGRA model, they can also be
negligible, as shown in Ref.~\cite{ctf02}. Therefore we here use
the same operator base as in the SM.

It is well known that there is no SUSY contributions to the Wilson
coefficients at the tree level. At the one-loop level,
there are four kinds of SUSY contributions to the quark level decay process
$ b\to qq^{'}\bar{q^{'}}$, depending on specific particles propagated in
the loops:
\begin{enumerate}
\item[(i)]
the charged Higgs boson $H^{\pm}$ and up-type quarks $u,c,t$;

\item[(ii)]
the charginos $\tilde{\chi}^{\pm}_{1,2}$ and the
scalar up-type quarks $\tilde{u}, \tilde{c},\tilde{t}$;

\item[(iii)]
the
neutralinos $\tilde{\chi}^{0}_{1,2,3,4}$ and the scalar down-type
quarks $\tilde{d}, \tilde{s},\tilde{b}$;

\item[(iv)]
the guinos $\tilde{g}$ and the scalar down-type quarks $\tilde{d},
\tilde{s},\tilde{b}$.
\end{enumerate}
In Ref.~\cite{zw04}, we have given a detailed derivation of the
lengthy expressions of SUSY corrections induced by those penguin
diagrams involving SUSY particles to the Wilson coefficients.

In the mSUGRA model, there are five free parameters: $\tan\beta, m_{\frac{1}{2}},
m_{0}, A_{0}$ and $sign(\mu)$.
Before we calculate the new physics contributions to the considered $B\to PV, VV$
decays, we have to check the experimental bounds on these parameters, which have been
studied extensively by many authors \cite{lepa,lepb,spa,sps1}.
The supersymmetry parameter
analysis(SPA) project with the main target of high-precision determination of
the supersymmetry Lagrangian parameters at the electroweak scale, for example,
is under way  now \cite{spa}.
By confronting the precision electroweak data provided by the LEP experiments
with the theoretical predictions within the mSUGRA model, the strong constraints
on the parameter space have been obtained \cite{lepa,lepb,spa,sps1}.

From the well measured $B \to X_s \gamma$ decays, the magnitude-but not the sign- of
the Wilson coefficient $C_{7\gamma}(m_b)$ is strongly constrained.
This is an important constraint. However, various new physics can change
the sign of $C_{7\gamma}(m_b)$ without changing the branching ratio of
$B \to X_s \gamma$ decay obviously.

From the measurements of the $b \to s l^+ l^-$ decays, one can determine the relative
sign of the Wilson coefficients as well as their absolute values.
The latest Belle and BaBar measurements of the inclusive $B \to X_s l^+ l^-$
branching ratios indicated that the sign of $C_{7\gamma}(m_b)$ is unlikely
to be different from that in the SM \cite{pum05}.
On the other hand, the studies \cite{pum05,ali02} also show that a positive
$C_{7\gamma}^{eff}$ could be made compatible with experiments only
by large ${\cal O}(1)$ new physics corrections to $C_{9,10}^{eff}$, while the SM values
of $C_{9,10}^{eff}$ are around $+4.2$ and $-4.4$ respectively.

For the semi-leptonic Wilson coefficients $C_9$ and $C_{10}$ in the minimal
supergravity model, the authors of Ref.~\cite{tyy97} have made a detailed analysis 
and found that $C_9(m_b)$ and $C_{10}(m_b)$ differ from their SM values by at most $5\%$ 
in the parameter space for $-2\leq C_{7\gamma}(m_b)/C_{7\gamma}^{SM}(m_b) \leq 2$ 
and for $\tan{\beta}=3,30$. This result supports a SM-like $C_{7\gamma}(m_b)$ in 
the mSUGRA model. 

This August, Belle Collaboration reported their measurement of the ratios of Wilson
coefficients $A_9/A_7$ and $A_{10}/A_7$
\footnote{ At next-next-to-leading order (NNLO), the effective
Wilson coefficients $C_{7\gamma}^{eff}, C_9^{eff}$
and $C_{10}^{eff}$ have many small high order correction terms \cite{aagw01},
one usually use the leading coefficients
$A_{7}$, $A_9$ and $A_{10}$ in the evaluations \cite{belle05}.
The numerical differences between
$C_{7\gamma,9,10}^{eff}$ and $A_{7,9,10}$ are indeed very small.
It is worth noting that, instead of the
(electroweak penguin) coefficients $C_9$ and $C_{10}$ as defined in Eq.(\ref{eq:eff}),
the coefficients
$C_9^{eff}$ and $C_{10}^{eff}$ here correspond to the low-energy interaction terms
$(\bar{s}_L \gamma_\mu b_L) (\bar{l}\gamma^\mu l )$ and
$(\bar{s}_L \gamma_\mu b_L) (\bar{l}\gamma^\mu \gamma_5 \, l )$, respectively.}
in $B \to K^* l^+ l^-$ decay \cite{belle05}.
They exclude a positive $A_9 A_{10}$ at more than
$95\%$ CL., but can not determine the sign of $A_7$ (i.e. $C_{7\gamma}(m_b)$).

Based on a lot of previous studies about the constraints on the parameter space of
the mSUGRA model \cite{lepa,lepb,spa,sps1,zw04,ali02},
we can choose several typical sets of input parameters to show the
pattern of the new physics corrections to the branching ratios
of the studied decays in the mSUGRA model.
In Ref.~\cite{zw04} we selected two sets of input parameters (case-A and case-B)
and found that (a) the SUSY corrections to the Wilson coefficients $C_k
(k=3\sim6)$ are always very small and can be neglected safely;
(b) with the inclusion of SUSY corrections, the Wilson coefficient $C_{7\gamma}(m_b)$
remained SM-like for case-A, but changed its sign for the case-B.
Numerical results for the branching ratios of $B \to PP$ decays also show that
the new physics corrections are very small for the case-A, but can be significant
for the case-B.

In this paper, besides the choice of case-A and B as defined in Ref.\cite{zw04},
we also consider the third case, the case-C, which is the ``typical"
mSUGRA point ``SPS 1b" as defined in Ref.\cite{sps1}.
In the case-C, the SUSY corrections to $C_k (k=3\sim6)$ are also negligibly small,
while $C_{7\gamma}(m_b)$ is SM-like after the inclusion of SUSY contributions.
The three sets of mSUGRA input
parameters to be used in numerical calculations here are listed in Table \ref{tab1},
the numerical values of the ratio $R_7=C_{7\gamma}(m_b)/C_{7\gamma}^{SM}(m_b)$ for
each cases are also given
in the table. It is easy to see that the Wilson coefficient $C_{7\gamma}$ in the mSUGRA
model is SM-like (negative) for both case-A and C, but nonstandard (positive) for
case-B. The ratio $R_7 \gtrsim 1$ for case-A, but $\lesssim 1$ for the case-C, due to
the strong bound imposed by the precision data of $B \to X_s \gamma$.

\begin{table}[tbp]
\caption{ Three typical sets of SUSY parameters to be used in numerical calculation.
The third case is the ``typical" mSUGRA point ``SPS 1b" as defined
in Ref.\cite{sps1}. The last column show the values of the ratio
$R_7=C_{7\gamma}(m_b)/C_{7\gamma}^{SM}(m_b)$. All masses are in units of $GeV$.}
\label{tab1}
\begin{center}
\begin{tabular}{c|cccccccccc|ccccc} \hline  \hline
\multicolumn{1}{c|}{CASE}&
\multicolumn{2}{c}{$m_0$\qquad }&\multicolumn{2}{c}{$m_{\frac{1}{2}}$\qquad }&
\multicolumn{2}{c}{$A_0$\qquad }&\multicolumn{2}{c}{$\tan\beta$}&
\multicolumn{2}{c|}{$Sign[\mu]$}&
\multicolumn{5}{c}{$ R_7$} \\
\hline\hline
A&\multicolumn{2}{c}{$300$}&\multicolumn{2}{c}{$300$}&
\multicolumn{2}{c}{$0$}&\multicolumn{2}{c}{$2$}&\multicolumn{2}{c|}{$-$}&
\multicolumn{5}{c}{$1.10$} \\
B&\multicolumn{2}{c}{$369$}&\multicolumn{2}{c}{$150$}&
\multicolumn{2}{c}{$-400$}&\multicolumn{2}{c}{$40$}&\multicolumn{2}{c|}{$+$}&
\multicolumn{5}{c}{$-0.93$} \\
C&\multicolumn{2}{c}{$200$}&\multicolumn{2}{c}{$400$}&
\multicolumn{2}{c}{$0$}&\multicolumn{2}{c}{$30$}&\multicolumn{2}{c|}{$+$}&
\multicolumn{5}{c}{$0.82$}\\
\hline \hline
\end{tabular}
\end{center}
\end{table}

From above discussions, one expects that the case-A and C will be very similar
phenomenologically. The numerical calculations to be given in the following sections
indeed show that the  theoretical predictions for the branching ratios
in case-A and C are almost identical with each other:
the difference is less than $2\%$.
We therefore will present the numerical results of the branching ratios
explicitly for case-B and case-C only, and compare them with the corresponding
SM predictions.

\section{$B\to M_1M_2$ decays in the QCD factorization approach}\label{sec:qcdf}

In QCD factorization approach, when the final state hadrons of B meson two-body decays
are all light mesons (mesons composed of light $u,d,$ or $s$ quarks, and with a
mass of order $\Lambda_{QCD}$), the matrix element of each operator in the
effective Hamiltonian ${\cal H}_{eff}$ can be written as \cite{bn03b}
\begin{eqnarray}
\langle M_{1}M_{2}|O_i|B\rangle&=&\sum_{j}F_{j}^{B\rightarrow
M_{1}}\int_{0}^{1}dx
T_{ij}^{I}(x)\Phi_{M_2}(x)+(M_1\leftrightarrow M_2) \nonumber \\
&   &+\int_{0}^{1}d\xi\int_{0}^{1}dx\int_{0}^{1}dy
T_{i}^{II}(\xi,x,y)\Phi_{B}(\xi)\Phi_{M_1}(x)\Phi_{M_2}(y)
\label{eq:QCDF}
\end{eqnarray}
where  $F_{j}^{B\rightarrow M_{1}}$ is the form factor describing
$B\to M_1$ decays,  $T_{ij}^{I}$ and $T_{i}^{II}$ denote the
perturbative short-distance interactions and can be calculated by
the perturbation approach, and $\Phi_{X}(x)(X=B,M_{1,2})$ are the
universal and nonperturbative light-cone distribution amplitudes
(LCDA) for the heavy B meson and the light $M_{1,2}$ meson respectively.
Weak annihilation effects which are suppressed by $\Lambda_{QCD}/m_b$ are not
included in Eq.(\ref{eq:QCDF}).

Considering the low energy effective Hamiltonian Eq.(\ref{eq:eff})
and the QCDF formula Eq.(\ref{eq:QCDF}), the decay amplitude can
be written as
\beq \label{eq:af} {\cal A}^f(B\to
M_{1}M_{2})=\frac{G_F}{\sqrt{2}}\sum_{p=u,c}\sum_{i}V_{pb}V_{pq}^{*}a_i^p(\mu)\langle
M_{1}M_{2}|O_i|B\rangle _{F} .
\eeq
Here $\langle M_{1}M_{2}|O_i|B\rangle _{F} $ is the factorized matrix element.
The explicit expressions for the decay amplitudes of $B\to
M_{1}M_{2}$ decays can be found for example in the Appendixes of
Refs.~\cite{agc98,yhb99}.  Following Beneke {\sl et al.}
\cite{bn03b}, the coefficients $a_i(M_1M_2)$($i=1$ to 10) in
Eq.(\ref{eq:af}) with $M_1$ absorbing the spectator quark is
\beq
\label{eq:ai} a_i^p (M_1 M_2 ) &=&(C_i  + \frac{{C_{i \pm 1}
}}{{N_c }})N_i (M_2 ) \nonumber\\ && + \frac{{C_{i \pm 1} }}{{N_c
}}\frac{{C_F \alpha _s }}{{4\pi }}\left[V_i (M_2 ) + \frac{{4\pi
^2 }}{{N_c }}H_i (M_1 M_2 )\right] + P_i^p (M_2 ),
\eeq
where the
upper (lower) signs apply when i is odd (even). The functions
$V_i(M_2)$ account for one loop vertex corrections, $H_i (M_1
M_2)$ for hard spectator interactions and $P_i^p (M_2 )$ for
penguin contributions. The explicit expressions for these
functions can be found in Ref.~\cite{bn03b}.

As mentioned previously,  the SUSY contributions to the Wilson
coefficients of the four-quark penguin operators are very small
and have been neglected. The new magnetic penguin contributions in
the mSUGRA model can manifest themselves as radiative corrections to
the coefficients $a_{i}^p$ and be contained in the functions
$P_i^p (M_2 )$, which is present only for $i=4,6,8,10$
\cite{bn03b}. To make this visible, here we just show the
functions $P_4^p (M_2 )$ and $P_{10}^p (M_2 )$. At order
$\alpha_s$, these two functions are
\begin{eqnarray}
\label{eq:p4}
P_4^p (M_2 ) &=& \frac{C_F \alpha _s }{4\pi N_c } \left\{ C_1
\left[ \frac{4}{3}\ln \frac{m_b}{\mu } + \frac{2}{3} - G_{M_2 }
(s_p )\right] \right.\non
 &  & + C_3 \left[ \frac{8}{3}\ln \frac{m_b }{\mu }
+ \frac{4}{3} - G_{M_2 } (0) - G_{M_2 } (1)\right]  \non
 &  & + (C_4  + C_6
)\left[ \frac{4n_f }{3}\ln \frac{m_b }{\mu } - (n_f  - 2)G_{M_2 }
(0) - G_{M_2 } (s_c ) - G_{M_2 } (1)\right] \non
 &  &  \left. -
2C_{8g}^{eff} \int\limits_0^1 {\frac{dx}{1-x}\Phi _{M_2 } (x)}
\right \} ,\\
\label{eq:p10}
P_{10}^p (M_2 ) &=& \frac{ \alpha }{9\pi N_c
}\left\{ (C_1+N_c C_2 )\left[ \frac{4}{3}\ln \frac{m_b}{\mu } +
\frac{2}{3} - G_{M_2 } (s_p )\right]\right.\non  & & \left.  -
3C_{7\gamma}^{eff} \int\limits_0^1 {\frac{dx}{1-x}\Phi _{M_2 }
(x)} \right \},
\end{eqnarray}
where $s_{u}=m_u^2/m_b^2\approx 0$ and $s_{c}=m_{c}^2/m_{b}^2$ are
mass ratios involved in the evaluation of penguin diagrams. $\Phi
_{M_2 } (x)$ is the leading twist LCDA which can be expanded in
Gegenbauer polynomials. $ C_{7\gamma }^{eff} = C_{7\gamma } -
\frac{1}{3}C_5 - C_6$ and $C_{8g}^{eff} = C_{8g} + C_5 $ are the
so-called ``effective " Wilson coefficients, where the SUSY
contribution is involved. The explicit expressions of the
functions $G_{M_2 }(0)$, $G_{M_2 }(1)$ and $G_{M_2 }(s_p)$
appeared in Eqs.(\ref{eq:p4},\ref{eq:p10}) can be found easily in
Refs.~\cite{bbns01,bn03b}.

When calculating the decay amplitudes, the coefficients $a_i(i=3\sim10)$
always appear in pairs. So in terms of the coefficients $a_i^p$,
one can define $\alpha_i^p$ as follows \cite{bn03b}:
\begin{eqnarray}\label{eq:ais}
   \alpha_1(M_1 M_2) &=& a_1(M_1 M_2) \,, \nonumber\\
   \alpha_2(M_1 M_2) &=& a_2(M_1 M_2) \,, \nonumber\\
   \alpha_3^p(M_1 M_2) &=& \left\{
    \begin{array}{cl}
     a_3^p(M_1 M_2) - a_5^p(M_1 M_2) \,
      & \quad \mbox{if~} M_1 M_2=PP, \,VP \,, \\
     a_3^p(M_1 M_2) + a_5^p(M_1 M_2) \,
      & \quad \mbox{if~} M_1 M_2=PV  \,,
    \end{array}\right. \nonumber\\
   \alpha_4^p(M_1 M_2) &=& \left\{
    \begin{array}{cl}
     a_4^p(M_1 M_2) + r_{\chi}^{M_2}\,a_6^p(M_1 M_2) \,
      & \quad \mbox{if~} M_1 M_2=PP, \,PV \,, \\
     a_4^p(M_1 M_2) - r_{\chi}^{M_2}\,a_6^p(M_1 M_2) \,
      & \quad \mbox{if~} M_1 M_2=VP\,,
    \end{array}\right.\\
   \alpha_{3,\rm EW}^p(M_1 M_2) &=& \left\{
    \begin{array}{cl}
     a_9^p(M_1 M_2) - a_7^p(M_1 M_2) \,
      & \quad \mbox{if~} M_1 M_2=PP, \,VP \,, \\
     a_9^p(M_1 M_2) + a_7^p(M_1 M_2) \,
      & \quad \mbox{if~} M_1 M_2=PV  \,,
    \end{array}\right. \nonumber\\
   \alpha_{4,\rm EW}^p(M_1 M_2) &=& \left\{
    \begin{array}{cl}
     a_{10}^p(M_1 M_2) + r_{\chi}^{M_2}\,a_8^p(M_1 M_2) \,
      & \quad \mbox{if~} M_1 M_2=PP, \,PV \,, \\
     a_{10}^p(M_1 M_2) - r_{\chi}^{M_2}\,a_8^p(M_1 M_2) \,
      & \quad \mbox{if~} M_1 M_2=VP\,.
     \end{array}\right.\nonumber
\end{eqnarray}
For pseudoscalar meson P and vector meson V, the ratios $r_\chi^{P}$ and
$r_\chi^V(\mu) $ are defined as
\beq
\label{eq:rchi}
   r_\chi^P(\mu) &=& \frac{2m_P^2}{m_b(\mu)(m_{q_1}+m_{q_2})(\mu)}, \\
\label{eq:rchiV}
   r_\chi^V(\mu) &=& \frac{2m_V}{m_b(\mu)}\,\frac{f_V^\perp(\mu)}{f_V} \,,
\eeq where  $m_{q_1}$ and $m_{q_2}$ are the current masses of the
component quarks of P meson, and  $f_V^\perp(\mu)$ is the
scale-dependent transverse decay constant of vector meson V.
Although all the terms  proportional to $r_\chi^{M_2}$ are
formally suppressed by one power of $\Lambda_{\rm QCD}/m_b$ in the
heavy-quark limit, these terms are chirality enhanced and not
always small. They are very important in those penguin-dominant B
meson decays, such as the interesting channels $B\to
K\eta^{'}$,etc.

In QCD factorization approach, the nonfactorizable
power-suppressed contributions are neglected. However, the
hard-scattering spectator interactions and annihilation diagrams
can not be neglected because of the chiral enhancement. Since they
give rise to infrared endpoint singularities when computed
perturbatively, they can only be estimated in a model dependent
way and with a large uncertainty. In Refs.~\cite{bbns01,bn03b}
these contributions are parameterized by two complex quantities,
$\chi_H$ and $\chi_A$, \beq \chi_{H,A}= \left ( 1+ \rho_{H,A} e^{i
\phi_{H,A}}\right ) \ln\frac{m_B}{\Lambda_h} \label{eq:xha} \eeq
where $\Lambda_h=0.5 GeV$, $\phi_{H,A}$ are free phases in the
range $[-180^\circ, 180^\circ]$, and $\rho_{H,A}$ are real
parameters varying within $[0,1]$. In this paper, we use the same
method as in Refs.~\cite{bbns01,bn03b} to estimate these two kinds
of contributions.

As given in Ref.~\cite{bn03b},  the annihilation amplitude can be
written as
\beq
\label{eq:ann}
{\cal A}^{ann}(B\to
M_{1}M_{2})\propto\frac{G_F}{\sqrt{2}}\sum_{p=u,c}\sum_{i}V_{pb}V_{pq}^{*}
f_Bf_{M_1}f_{M_2}b_i(M_{1}M_{2})
\eeq
where $f_B$ and $f_M$ are the decay constants of B meson and
final-state hadrons respectively. The coefficients
$b_i(M_{1}M_{2})$ describe the annihilation contributions. For
explicit expressions of coefficients $b_i$, one can see
Refs.~\cite{bbns01,bn03b}.

For $B\to VV$ decays  we have two additional remarks: (a) since
$<V|\bar{q}_1q_2|0>=0$, $B\to VV$ decays do not receive
factorizable contribution  from $a_6$ and $a_8$ penguin terms
except for spacelike penguin diagrams; and (b) unlike the PP and
PV decay modes, the annihilation amplitude in the VV mode does not
have  a chiral enhancement of order $M_B^2/(m_qm_b)$. Therefore,
it is truly power suppressed in heavy quark limit and will be
neglected in our calculation. The explicit factorizable
coefficients $a_i$ for $B\to VV$ can be found in
Ref.~\cite{vvqcd}.


\section{Branching ratios of $B\to PV$ decays}\label{sec:pv}

In the following two sections we will calculate the CP-averaged branching ratios
for thirty nine $B\to PV$  and nineteen $B \to VV$ decay modes, respectively.
We usually use the central values of the input parameters as collected in
Appendix A, and consider the effects of the uncertainties of these input parameters
as specified in the text for individual decay channels.

Theoretically, the branching ratios of charmless decays $B{\to}PV$
in the B meson  rest frame can be written as
 \beq
 {\cal B}r(B{\to}PV)= \frac{{\tau}_{B}}{8{\pi}}
  \frac{{\vert}P_c{\vert}}{m_{B}^{2}}{\vert}{\cal A}(B{\to}PV){\vert}^{2},
 \eeq
where $\tau_B$ is the B meson lifetimes, and $ |P_c|$ is the
absolute values of final-state hadrons' momentum in the B rest
frame and written as \beq
 {\vert}P_c{\vert}=\frac{\sqrt{[m_{B}^{2}-(m_{P}+m_{V})^{2}]
       [m_{B}^{2}-(m_{P}-m_{V})^{2}]}}{2m_{B}}.
\eeq For the CP-conjugated decay modes, the branching ratios  can
be obtained by replacing CKM factors with their complex conjugate
in the expressions of decay amplitudes. Using the decay amplitudes
as given in Refs.~\cite{agc98,bn03b} and the coefficients $a_i$ in
Eq.(\ref{eq:ai}) or $\alpha_i$ in Eq.(\ref{eq:ais}), it is
straightforward to calculate the CP-averaged branching ratios of
those thirty nine $B \to PV$ decay modes in the SM and mSUGRA model.

From Eqs.(\ref{eq:ai}-\ref{eq:ais}), we can find that the potential
SUSY contributions  are mainly embodied in $\alpha_4^p(M_1M_2)$
and $\alpha_{4,ew}^p(M_1M_2)$. Therefore, one naturally expect a
 large new physics corrections to those penguin
dominated B meson decays.

\subsection{Numerical results}

In Table \ref{smmssm1} and \ref{smmssm2}, we show the theoretical
predictions for the CP-averaged branching ratios for $B\to PV$
decays in both the SM and  mSUGRA model, assuming $\mu=m_b/2, m_b$
and $2 m_b$, respectively. And in the SM  we give both $Br^{f+a}$ and
$Br^{f}$, the SM predictions for the branching ratios with or without the
inclusion of annihilation contributions, respectively. In the
mSUGRA model, we consider both case B and case C and only give
$Br^{f}$ so that we can compare the relative size of the new physics  
contribution with the annihilation contribution for each considered decay channel.
From the numerical results, one can see that
the SUSY corrections to the $b\to s$
transition processes are generally larger than those to the $b\to d$
processes because of the CKM factor suppression ($|V_{tb} V_{td}^*|
\sim 10^{-2}$) in $b\to d$ penguin transition.

\begin{table}[tbp]
\doublerulesep 1.5pt\caption{Numerical predictions in the  SM and
mSUGRA model for CP-averaged branching ratios ( in units of $10^{-6}$)
for $b\to d$ transition processes of $B\to PV$ decays, where
$Br^{f+a}$ and $Br^f$ denotes the branching ratios with and
without the annihilation contributions respectively.
For Case B and C, the branching ratios without
annihilation contributions are given. }\label{smmssm1}
\begin{center}
\begin{tabular} {l|cc|cc|cc|cc|cc|cc} \hline  \hline
\multicolumn{1}{c|}{$B\to PV$} &
\multicolumn{4}{|c|}{$\mu=m_{b}/2$} &
\multicolumn{4}{|c|}{$\mu=m_{b}$}&
\multicolumn{4}{|c}{$\mu=2m_{b}$}
 \\ \cline{2-13}
\   \ \  \ $(b\to d)$ & \multicolumn{2}{|c|}{SM}&
\multicolumn{2}{|c|}{mSUGRA} &
       \multicolumn{2}{|c|}{SM}& \multicolumn{2}{|c|}{mSUGRA} &
      \multicolumn{2}{|c|}{SM}& \multicolumn{2}{|c}{mSUGRA}
  \\ \cline{2-13}
\ \ &$Br^f$&$Br^{f+a}$ & (B) &(C)& $Br^f$&$Br^{f+a}$
    & (B) &(C)&$Br^f$&$Br^{f+a}$ & (B) &(C)
\\ \hline \hline
$B^-\to\pi^-\rho^0$&11.2&11.2&11.2&11.2&11.5&11.4&11.5&11.5&11.8&11.8&11.8&11.8
\\
$B^-\to\pi^0\rho^-$&14.8&14.9&14.9&14.8&14.9&15.0&15.0&14.9&15.1&15.2&15.2&15.1
\\
$\bar{B}^0\to\pi^+\rho^-$&21.2&22.3&21.5&21.2&21.2&22.1&21.4&22.2&20.9&21.7&21.2&20.9
\\
$\bar{B}^0\to\pi^-\rho^+$&14.3&15.1&14.4&14.3&14.3&14.9&14.3&14.3&14.1&14.7&14.2&14.1
\\
$B^{-}\to\pi^{-}\omega$&9.09&8.70&9.24&9.09&9.25&8.98&9.38&9.25&9.48&9.29&9.59&9.48
\\
 $B^-\to
 \eta\rho^-$&6.50&6.19&6.61&6.50&6.55&6.35&6.64&6.55&6.67&6.52&6.75&6.67
\\
$B^-\to\eta^{'}\rho^-$&4.60&4.39&4.67&4.60&4.66&4.51&4.71&4.66&4.78&4.67&4.82&4.78
\\ \hline\hline
$\bar{B}^0\to\pi^0\rho^0$&0.538&0.417&0.532&0.538&0.524&0.419&0.512&0.523&0.602&0.502&0.585&0.601
\\
$\bar{B}^{0}\to\pi^{0}\omega$&0.019&0.013&0.043&0.019&0.014&0.007&0.032&0.014&0.012&0.004&0.025&0.012
\\
$\bar{B}^0\to\eta\rho^0$&0.004&0.021&0.010&0.004&0.003&0.016&0.006&0.003&0.003&0.014&0.004&0.003
\\
$\bar{B}^0\to\eta^{'}\rho^0$&0.035&0.066&0.033&0.035&0.033&0.058&0.029&0.033&0.034&0.055&0.029&0.034
\\
$\bar{B}^{0}\to\eta\omega$&0.278&0.351&0.295&0.278&0.249&0.308&0.262&0.249&0.269&0.323&0.279&0.269
\\
$\bar{B}^{0}\to\eta^{'}\omega$&0.274&0.337&0.283&0.274&0.253&0.305&0.259&0.253&0.276&0.323&0.281&0.276
\\ \hline\hline
$B^{-}\to\pi^{-}\phi$&0.008&$-$&0.008&0.008&0.006&$-$&0.006&0.006&0.005&$-$&0.005&0.005
\\
$\bar{B}^{0}\to\pi^{0}\phi$&0.003&$-$&0.003&0.003&0.002&$-$&0.002&0.002&0.002&$-$&0.002&0.002
\\
$\bar{B}^{0}\to\eta\phi$&0.002&0.001&0.002&0.002&0.001&0.001&0.001&0.001&0.001&0.0008&0.001&0.001
\\
$\bar{B}^{0}\to\eta^{'}\phi$&0.002&0.003&0.002&0.002&0.001&0.002&0.001&0.001&0.001&0.0011&0.001&0.001
\\ \hline\hline
 $B^-\to K^-K^{*0}$&0.12&0.17&0.31&0.12&0.11&0.15&0.28&0.11&0.10&0.12&0.24&0.10
\\
 $\bar{B}^0\to
 \bar{K}^0K^{*0}$&0.11&0.14&0.29&0.11&0.10&0.13&0.26&0.11&0.09&0.11&0.22&0.10
\\ \hline\hline
 $B^-\to
 K^0K^{*-}$&0.10&0.16&0.03&0.10&0.11&0.16&0.04&0.11&0.13&0.17&0.07&0.13
\\
$\bar{B}^0\to
K^0\bar{K}^{*0}$&0.09&0.15&0.02&0.09&0.10&0.15&0.04&0.10&0.12&0.16&0.06&0.12
\\ \hline\hline
 $\bar{B}^0\to K^+K^{*-}$&$-$&0.02&$-$&$-$&$-$&0.01&$-$&$-$&$-$&0.01&$-$&$-$
\\
 $\bar{B}^0\to K^-K^{*+}$&$-$&0.02&$-$&$-$&$-$&0.01&$-$&$-$&$-$&0.01&$-$&$-$
\\ \hline\hline
\end{tabular}
\end{center}
\end{table}

\begin{table}[tbp]
\doublerulesep 1.5pt\caption{Numerical predictions in the  SM and
mSUGRA model for CP-averaged branching ratios (in units of $10^{-6}$)
for $b\to s$ transition process of $B\to PV$ decays, where
$Br^{f+a}$ and $Br^f$ denotes the branching ratios with and
without the annihilation contributions respectively.
For Case B and C, the branching ratios without
annihilation contributions are given. }\label{smmssm2}
\begin{center}
\begin{tabular} {l|cc|cc|cc|cc|cc|cc} \hline  \hline
\multicolumn{1}{c|}{$B\to PV$} &
\multicolumn{4}{|c|}{$\mu=m_{b}/2$} &
\multicolumn{4}{|c|}{$\mu=m_{b}$}&
\multicolumn{4}{|c}{$\mu=2m_{b}$}
 \\ \cline{2-13}
\   \ \  \ $(b\to s)$ & \multicolumn{2}{|c|}{SM}&
\multicolumn{2}{|c|}{mSUGRA} &
       \multicolumn{2}{|c|}{SM}& \multicolumn{2}{|c|}{mSUGRA} &
      \multicolumn{2}{|c|}{SM}& \multicolumn{2}{|c}{mSUGRA}
  \\ \cline{2-13}
\ \ &$Br^f$&$Br^{f+a}$ & (B) & (C) & $Br^f$&$Br^{f+a}$
    & (B) & (C)& $Br^f$ & $Br^{f+a}$  & (B) & (C)
\\ \hline \hline
$B^-\to\pi^-K^{*0}$&2.19&3.17&5.85&2.21&2.08&2.83&5.26&2.10&1.82&2.39&4.57&1.83
\\
$B^-\to\pi^0K^{*-}$&2.00&2.37&4.15&2.02&1.94&2.23&3.85&1.96&1.81&2.03&3.50&1.82
\\
$\bar{B}^0\to\pi^0\bar{K^{*0}}$&0.33&0.49&1.45&0.33&0.30&0.42&1.24&0.30&0.24&0.33&1.03&0.24
\\
$\bar{B}^0\to\pi^+K^{*-}$&1.68&2.27&4.38&1.70&1.62&2.07&3.99&1.64&1.49&1.82&3.55&1.50
\\
 $B^{-}\to
 K^{-}\phi$&2.73&4.08&6.06&2.75&2.46&3.47&5.31&2.47&2.04&2.79&4.49&2.05
\\
 $\bar{B}^{0}\to \bar{K}^{0}\phi$&2.53&3.66&5.60&2.55&2.27&3.12&4.90&2.28&1.89&2.52&4.15&1.90
\\ \hline\hline
  $B^-\to  K^-\rho^0$&1.24&1.70&0.72&1.23&1.39&1.77&0.87&1.38&1.66&2.01&1.12&1.65
\\
$B^-\to\bar{K}^0\rho^-$&1.92&3.08&0.55&1.91&2.14&3.07&0.89&2.13&2.58&3.29&1.39&2.57
\\
 $\bar{B}^0\to
 K^-\rho^+$&4.05&5.61&2.11&4.02&4.38&5.64&2.64&4.35&4.91&5.98&3.32&4.88
\\
 $\bar{B}^0\to \bar{K}^0\rho^0$&2.22&3.10&1.11&2.20&2.32&3.02&1.37&2.31&2.52&3.10&1.69&2.51
\\
 $B^{-}\to K^{-}\omega$&2.43&3.14&1.43&2.41&2.33&2.87&1.51&2.32&2.51&2.95&1.76&2.50
\\
 $\bar{B}^{0}\to
 \bar{K}^{0}\omega$&1.09&1.66&0.45&1.09&0.99&1.41&0.46&0.98&1.11&1.46&0.62&1.10
\\ \hline\hline
 $B^-\to\eta K^{*-}$&4.31 &5.64 & 4.68& 4.32  &4.64& 5.72 & 5.26& 4.65
 &5.18&6.08&6.06&5.20
\\
$\bar{B}^0\to\eta\bar{K}^{*0}$&4.58&5.98&4.86&4.58&4.94&6.07&5.45&4.94&5.46&6.40&6.20&5.46
\\
$B^-\to\eta^{'}K^{*-}$&1.86&2.71&0.93&1.84&2.13&2.95&0.83&2.11&2.51&3.25&1.12&2.49
\\
$\bar{B}^0\to\eta^{'}\bar{K}^{*0}$&1.21&1.99&0.61&1.20&1.40&2.17&0.43&1.38&1.72&2.42&0.65&1.71
\\ \hline\hline

\end{tabular}
\end{center}
\end{table}

We classify the thirty nine $B \to PV$ channels into $b\to d$  and $b\to s$
processes. For $b\to d$ processes, we have the following remarks
\begin{itemize}

\item
The $\bar{B}^0\to\pi^{\pm}\rho^{\mp}$ and
$B^-\to\pi^{0}\rho^{-}, \pi^{-}\rho^{0}, \pi^{-}\omega,
\eta^{(')}\rho^-$ decays.

These channels are tree-dominated decay modes and depend on the
large coefficient $\alpha_1$.  The SUSY corrections to these
decays are very small and can be neglected safely in all the
parameter space.

\item
The decays $\bar{B}^0\to\pi^{0}\rho^{0}, \pi^{0}\omega,
\eta^{(')}\rho^0, \eta^{(')} \omega$.

These channels have small branching ratios. The reason for this is
twofold. On the one hand, the tree contributions of these channels
are involved in the coefficient $\alpha_2 \sim 0.2$, which is far
smaller than the large coefficient $\alpha_1 \sim 1 $. The penguin
contributions, on the other hand,  are strongly suppressed by the
CKM factor $|V_{tb} V_{td}^*| \sim 10^{-2}$. From Table
\ref{smmssm1}, we can see that the SUSY corrections are always smaller
than the annihilation contributions, and thus  can be easily
masked by it.

\item
The decays $\bar{B}^0\to\pi^{0}\phi, \eta^{(')}\phi$ and $ B^-\to
\pi^{-}\phi $.

This kind of channels have both penguin and weak annihilation
contributions. But the penguin contributions come from the small
coefficients $\alpha_3^P$ and $\alpha_{3,ew}^P$. Therefore weak
annihilation contributions are  dominant for $\bar{B}^0\to
\eta^{(')}\phi$. The branching ratios of these decay are at the
${\cal O}(10^{-9})$ level, and the  SUSY corrections in all
parameter space can hardly affect them.

\item
The decays $\bar{B}^0\to\bar{K}^0K^{*0}$ and $ B^-\to K^{-}K^{*0} $.

These channels are penguin dominant decays. In their amplitudes,
the dominant term is proportional to $\alpha_4^P(P,V)$ and
$\alpha_{4,ew}^P(P,V)$ . The SUSY corrections therefore can
increase their branching ratios significantly in some parameter
space, such as in case B where the SUSY
contributions can provide a $130\%$ enhancement and are far larger
than the annihilation contributions. But in case C where
$C_{7\gamma}(m_b)$ is SM-like, the SUSY contributions are negligibly
small.

\item
The decays $\bar{B}^0\to K^0\bar{K}^{*0}$ and $ B^-\to K^{0}K^{*-} $.

Different from the $\bar{B}^0\to\bar{K}^0K^{*0}$ and $
B^-\to K^{-}K^{*0} $ decays, the dominant terms here are
proportional to $\alpha_4^P(V,P)$ and $\alpha_{4,ew}^P(V,P)$. The
SUSY corrections interfere destructively with their SM
counterparts and will decrease the  branching ratios by  about
$50\%$ in case B. For case C, the SUSY contributions are still
very small.

\item
The decays $\bar{B}^0\to K^{+}K^{*-}, K^{-}K^{*+}$.

These two channels have weak annihilation contributions only.
Therefore in the mSUGRA model their branching ratios can hardly be
affected.

\end{itemize}

For $b\to s$ transition processes, the tree contribution is suppressed by CKM
factors, the penguin contributions therefore play the major role.
These decays can be classified into three groups. And our remarks are

\begin{itemize}

\item $B\to\pi K^{*},  K\phi$ decays.

These decays are penguin dominant decays and both the annihilation
contributions and the SUSY contributions can give  enhancements to
their branching ratios. But for case C the SUSY contributions are
smaller than the annihilation contributions and can  be masked
easily. Only in case B where the enhancements can reach as large
as $100\%\sim 260\%$ for $B\to\pi K^{*}$ decay and about $100\%$
for $B\to K\phi$ decay.

\item
$B\to K \rho, K\omega$ decays.

In case B the branching ratios of
these decays will be decreased by $30\%\sim60\%$ after the
inclusion of SUSY corrections and theoretically we can isolate the
SUSY contributions from the annihilation contributions for which tend
to increase the branching ratios greatly.
In case C, however, the SUSY contributions are small, while the
annihilation contributions are relatively important for explaining
the current large measured data.

\item $B\to K^* \eta^{(')}$ decays.

For $B\to K^* \eta$ decay, their amplitudes strongly depend on
$\alpha_4^P(P,V)$ and $\alpha_{4,ew}^P(P,V)$. Although the
inclusion of SUSY corrections will increase their branching
ratios, the SUSY contributions will also be masked by
large annihilation contributions. For $B\to K^*
\eta^{'}$ decay, however, their amplitudes mainly depend on
$\alpha_4^P(V,P)$ and $\alpha_{4,ew}^P(V,P)$, the SUSY corrections
will trend to decrease their branching ratios in both cases and
can be separated from the annihilation contributions which can
contribute a $40\%\sim 60\%$ enhancement to the branching ratios.
\end{itemize}

From above discussions, one can see that the SUSY  contributions
in case-C (preferred by the measured branching ratio of $B \to X_s l^+ l^-$
decay \cite{pum05}) are always negligibly small.

\subsection{The data and phenomenological analysis}

Among the thirty nine $B \to PV$ decay modes considered
here, eighteen of them have been measured experimentally.
The latest individual measurements as reported by different groups and
the new world average for the branching ratios
can be found in the HFAG (Heavy Flavor Averaging Group) homepage \cite{hfag}.
In this subsection we will make a phenomenological discussion for those eighteen
measured decay channels.

\subsubsection{$B \to \pi \rho$ and $B \to \pi \omega$}

Among the $B\to \pi\rho$ and $\pi \omega$ decays, four of them
have been well measured, and the experimental upper limits are
available for the remaining $B \to \pi^0 \rho^0$ and $\pi^0
\omega$ decay modes. The data and the theoretical predictions for the four
measured decay modes (in units of $10^{-6}$) in the SM and  mSUGRA
model (both Case B and Case C) are
\begin{eqnarray}
Br(B^0 \to \pi^\pm  \rho^\mp)&=& \left \{\begin{array}{ll}
 24.0 \pm 2.5, & {\rm  Data}, \\
 37.0  ^{+5.4}_{-4.6} (A_0^{B\to\rho}) \;
 ^{+9.0}_{-7.4} (F_1^{B\to\pi})\;^{+3.1}_{-2.1}(\chi_A) \; \pm1.7(\gamma),
 & {\rm  SM}, \\
37.3  ^{+5.4}_{-4.6} (A_0^{B\to\rho}) \;
 ^{+9.1}_{-7.5} (F_1^{B\to\pi})\;^{+3.1}_{-2.1}(\chi_A) \;^{+1.9}_{-2.0}(\gamma),
 & {\rm  Case-B}, \\
37.0  ^{+5.4}_{-4.5} (A_0^{B\to\rho}) \;
 ^{+9.9}_{-7.5} (F_1^{B\to\pi})\;^{+3.0}_{-2.1}(\chi_A) \; ^{+1.7}_{-1.8}(\gamma),
 & {\rm   Case-C}, \\
\end{array} \right. \label{eq:pirho1} \\
Br(B^- \to \pi^- \rho^0)&=& \left \{\begin{array}{ll}
 8.7 ^{+1.0}_{-1.1}, & {\rm  Data}, \\
 11.4 ^{+3.4}_{-3.0} (A_0^{B\to\rho}) \; ^{+0.9}_{-0.7}(\gamma),
 & {\rm SM}, \\
 11.4 ^{+3.4}_{-3.0} (A_0^{B\to\rho}) \; ^{+1.0}_{-0.8}(\gamma),
 & {\rm Case-B}, \\
11.4 ^{+3.5}_{-2.9} (A_0^{B\to\rho}) \;^{+0.9}_{-0.6}(\gamma),
 & {\rm Case-C}, \\
\end{array} \right. \label{eq:pirho3} \\
Br(B^- \to \pi^0 \rho^-)&=& \left \{\begin{array}{ll}
 10.8 ^{+1.4}_{-1.5},& {\rm Data}, \\
 15.0^{+5.4}_{-4.6} (F_1^{B\to\pi}) \;^{+0.5}_{-0.2}(\chi_A)\;
  ^{+0.9}_{-1.1}(\gamma),
 & {\rm SM}, \\
15.1 ^{+5.4}_{-4.6} (F_1^{B\to\pi}) \;^{+0.5}_{-0.2}(\chi_A)\;
 ^{+1.2}_{-1.0}(\gamma),
 & {\rm Case-B}, \\
15.0^{+5.4}_{-4.5} (F_1^{B\to\pi}) \;^{+0.5}_{-0.2}(\chi_A)\;
  ^{+0.9}_{-1.1}(\gamma),
 & {\rm Case-C}, \\
\end{array} \right. \label{eq:pirho4}\\
Br(B^- \to \pi^- \omega)&=& \left \{\begin{array}{ll}
 6.6 \pm 0.6, & {\rm  Data}, \\
 9.0  ^{+2.5}_{-2.2} (A_0^{B\to\omega})
 \;  ^{+0.4}_{-0.5}(\chi_A), & {\rm SM}, \\
9.1 ^{+2.5}_{-2.2} (A_0^{B\to\omega}) \;
 ^{+0.4}_{-0.5}(\chi_A), & {\rm Case-B}, \\
9.0  ^{+2.5}_{-2.2} (A_0^{B\to\omega})
 \;  ^{+0.4}_{-0.5}(\chi_A),
 & {\rm Case-C},  \label{eq:piw} \\
\end{array} \right. \label{eq:pivarpi}
\end{eqnarray}
where the major errors are induced by the uncertainties of the
following input parameters: $F_1^{B\to \pi} = 0.28 \pm 0.05$,
$A_0^{B\to \rho} = 0.37 \pm 0.06$, $A_0^{B\to \omega} = 0.33 \pm
0.05$  and $ \gamma= 57.8^\circ \pm 20^\circ$. Throughout this paper 
we take the central values and the ranges of these input parameters 
specified here as the default values, unless explicitly stated otherwise.

We also set $\rho_A=0$ as the default input value \cite{bbns01} in numerical 
calculations, and scan over $\rho_A\in[0,1]$ and $\phi_A\in[-180^\circ, 180^\circ]$ 
to estimate the theoretical error induced by the uncertainty of annihilation 
contribution.

\begin{figure}[tbp]
\centerline{\epsfxsize=18cm\epsffile{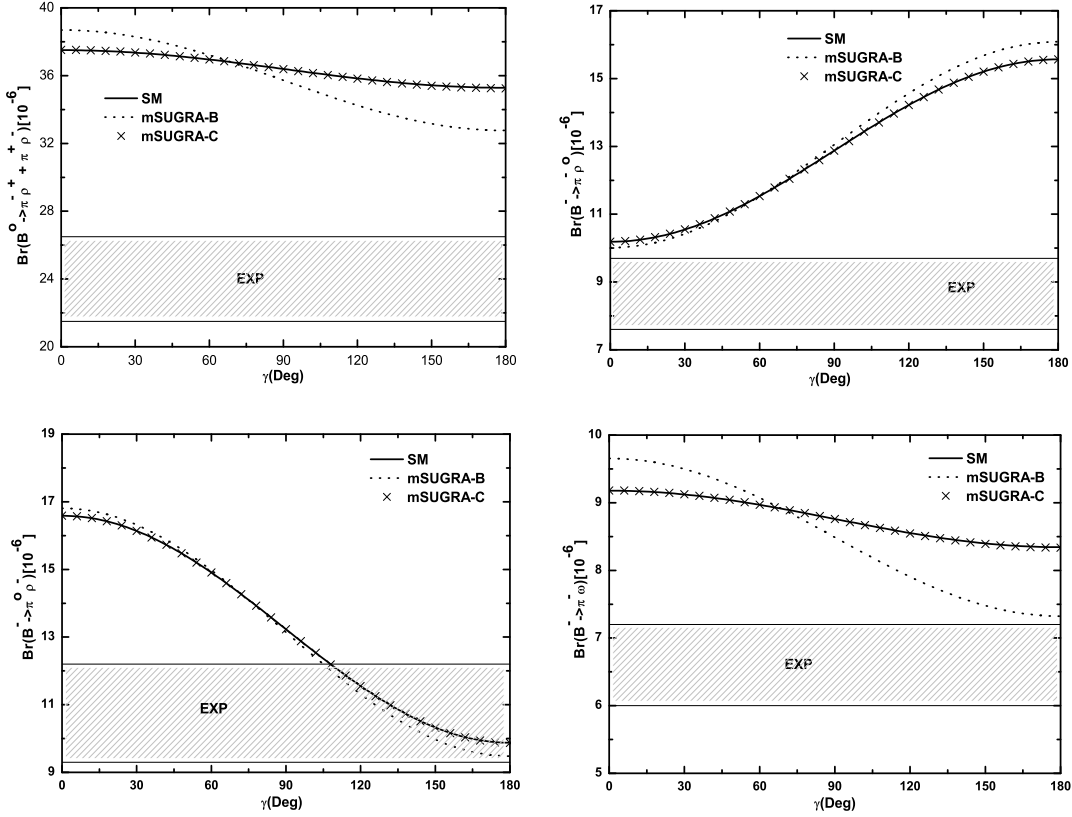}}
\vspace{-1cm}
\caption{The $\gamma$ dependence of the branching ratios of $B \to
\pi\rho$ and $B^-\to\pi^-\omega$ decays in the SM and  mSUGRA
model. The solid, dots lines and the scatter of prong show the
central values of the SM prediction and the mSUGRA ones  in both
case B and case C, respectively. The horizontal back-slashed gray bands show
the data. }
\label{fig:pirho}
\end{figure}

For the convenience of analysis,  in Fig.~\ref{fig:pirho} we show
the $\gamma$ dependence of the theoretical
predictions\footnote{The central values of all input parameters
except for the CKM angle $\gamma$ are used in this and other
similar figures. The theoretical uncertainties are not shown in
all such kinds of figures.} for the branching ratios of the three
$B \to \pi\rho$ and $B^-\to\pi^-\omega$ decays respectively and
the experimental data are also marked.

From the numerical results as given in
Eqs.(\ref{eq:pirho1}-\ref{eq:piw}) and Fig.~\ref{fig:pirho}, one
can see that
\begin{itemize}

\item
For these tree-dominated decay modes, the SUSY corrections
in the mSUGRA model are always very small for $\gamma \sim 60^\circ$.

\item 
As to the theoretical errors, generally the uncertainties of
the form factors and the annihilation contribution  are the dominant 
sources. For $B^0\to\pi^\pm \rho^\mp $, the annihilation contribution which has been
parameterized by $\chi_A$  also give large uncertainty. 
For $B^-\to\pi^- \rho^0$, however, the uncertainty of the annihilation contribution 
is small and has been ignored here.

\item 
The theoretical predictions for the branching ratios in both
the SM and mSUGRA model are all consistent with the data within
one standard deviation since the the theoretical errors are still large.

\end{itemize}

\subsubsection{$B\to \pi K^*$, $B\to K \phi$}

\begin{figure}[tbp]
\centerline{\epsfxsize=18cm\epsffile{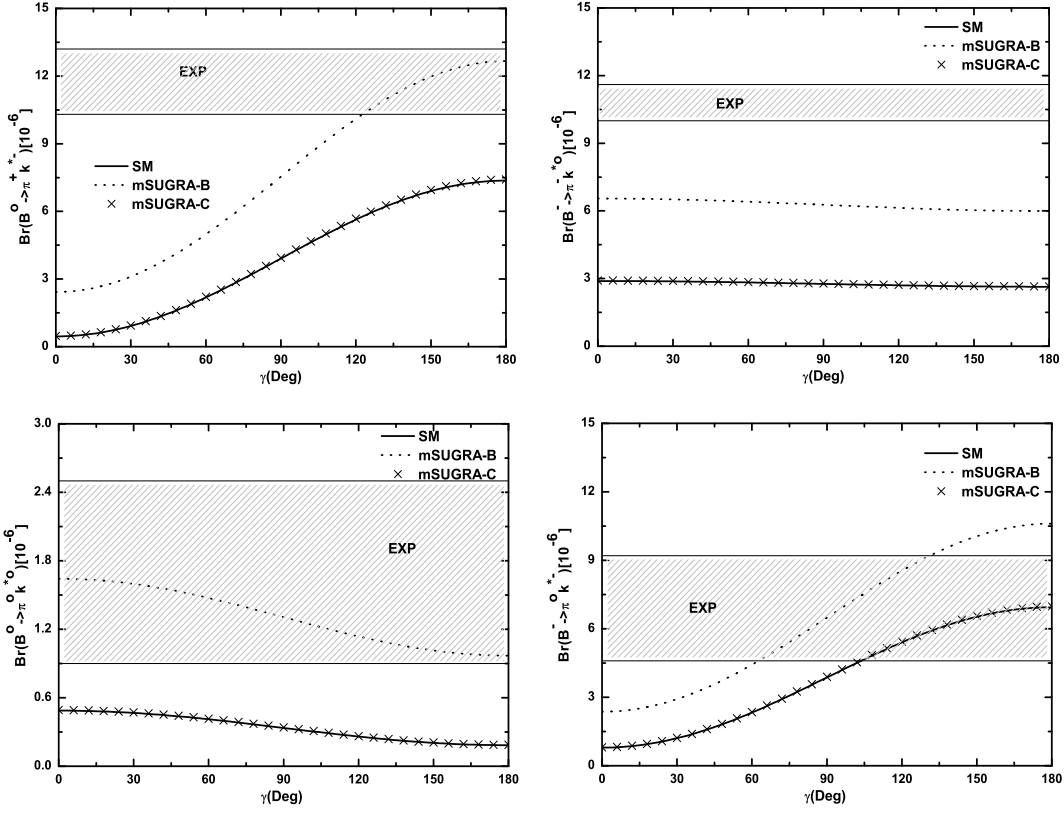}}
\vspace{-1cm}
\caption{The same as Fig.\ref{fig:pirho} but for $B \to \pi K^*$ decays.}
\label{fig:pikp}
\end{figure}

Like $B\to \pi K$, the four $B\to \pi K^*$ decays are penguin
dominated decay modes and therefore sensitive to the new physics
contributions. The CP-averaged branching ratios of the four $B \to \pi K^*$
decays have been measured
experimentally and the data are much larger than the SM
predictions. In the mSUGRA model, the new penguin diagrams may
contribute effectively to these decays. The data and the numerical results (in
units of $10^{-6}$)  are
\begin{eqnarray}
Br(\bar{B}^0 \to \pi^+ K^{*-})&=& \left \{\begin{array}{ll}
 11.7 ^{+1.5}_{-1.4}, & {\rm  Data}, \\
 2.1 ^{+0.2}_{-0.3}  (\mu) \; ^{+0.8}_{-0.6} (F_1^{B\to\pi}) \;
 ^{+4.4}_{-1.0}(\chi_A)\; ^{+1.1}_{-0.9}(\gamma),
 & {\rm  SM}, \\
 4.8 \pm0.7  (\mu) \; ^{+1.8}_{-1.5} (F_1^{B\to\pi})\;
 ^{+6.5}_{-2.1}(\chi_A)\;^{+1.7}_{-1.3}(\gamma),
 & {\rm  Case-B}, \\
2.1 ^{+0.2}_{-0.3}  (\mu) \; ^{+0.8}_{-0.6} (F_1^{B\to\pi}) \;
^{+4.4}_{-1.0}(\chi_A)\;^{+1.1}_{-0.9}(\gamma),
 & {\rm Case-C}, \\
\end{array} \right. \label{eq:pikp1} \\
Br(B^- \to \pi^- \bar{K}^{*0})&=& \left \{\begin{array}{ll}
 10.8 \pm 0.8 , & {\rm  Data}, \\
 2.8^{+0.3}_{-0.4} (\mu) \; ^{+1.1}_{-0.9} (F_1^{B\to\pi}) \;
  ^{+6.1}_{-1.8} (\chi_A),
 & {\rm  SM}, \\
6.4 \pm1.0  (\mu) \; ^{+2.5}_{-2.1} (F_1^{B\to\pi})
\;^{+8.3}_{-2.9} (\chi_A),
 & {\rm Case-B}, \\
2.9^{+0.3}_{-0.5} (\mu) \; ^{+1.1}_{-1.0}
(F_1^{B\to\pi})\;^{+6.1}_{-1.8}  (\chi_A),
 & {\rm Case-C}, \\
\end{array} \right. \label{eq:pikp2}
\eeq
\beq
Br(\bar{B}^0 \to \pi^0 \bar{K}^{*0})&=& \left \{\begin{array}{ll}
 1.7 \pm 0.8 , & {\rm  Data}, \\
 0.4 ^{+0.3}_{-0.2} (F_1^{B\to\pi}) \;
 \pm0.1 (A_0^{B\to K^*})\;^{+1.4}_{-0.3}  (\chi_A),  & {\rm \ \ SM}, \\
1.5 ^{+0.9}_{-0.7} (F_1^{B\to\pi}) \;\pm0.2
(A_0^{B\to K^*})\;^{+2.2}_{-0.7}  (\chi_A),
 & {\rm Case-B}, \\
 0.4 ^{+0.3}_{-0.2} (F_1^{B\to\pi})
\;\pm0.1 (A_0^{B\to K^*}
 )\;^{+1.4}_{-0.3}  (\chi_A),   & {\rm Case-C},\\
\end{array} \right. \label{eq:pikp3} \\
Br(B^- \to \pi^0 \bar{K}^{*-})&=& \left \{\begin{array}{ll}
 6.9 \pm 2.3 , & {\rm  Data}, \\
  2.2 ^{+0.6}_{-0.5} (F_1^{B\to\pi}) \;
  ^{+0.3}_{-0.2} (A_0^{B\to K^*}
 )\;^{+2.5}_{-0.8}  (\chi_A)\;^{+1.0}_{-0.8}(\gamma),  & {\rm \ \ SM}, \\
4.3 ^{+1.2}_{-1.1} (F_1^{B\to\pi}) \;^{+0.4}_{-0.3} (A_0^{B\to
K^*})\;^{+3.4}_{-1.3} (\chi_A)\;^{+1.3}_{-0.1}(\gamma),
 & {\rm Case-B}, \\
 2.3 ^{+0.6}_{-0.5} (F_1^{B\to\pi})
\;^{+0.3}_{-0.2}(A_0^{B\to K^*})\;^{+2.5}_{-0.8}  (\chi_A)\;
^{+1.0}_{-0.8}(\gamma),   & {\rm Case-C},\\
\end{array} \right. \label{eq:pikp4}
\end{eqnarray}
where the renormalization scale $\mu$ varies from $m_b/2$ to
$2m_b$, and the second error in Eq.(\ref{eq:pikp3}) and (\ref{eq:pikp4}) 
is induced by the uncertainty of the form factor 
$A_0^{B\to K^*}$, $A_0^{B\to K^*}=0.45\pm0.07$.

Fig.~\ref{fig:pikp} shows the $\gamma$ dependence of the branching
ratios for the measured $B\to \pi K^*$ decays in both the
SM and the mSUGRA model. From this figure and
Eqs.(\ref{eq:pikp1}-\ref{eq:pikp4}), one can see that
\begin{itemize}

\item 
The central values of the SM predictions for the branching
ratios are only about 20 to 30 percent of the measured values. The
SUSY contributions are also very small in the parameter space
where $C_{7\gamma}(m_b)$ is SM-like. But for the case B, the SUSY
contributions are large and can provide a factor of two
enhancements to these penguin-dominated decays. After the
inclusion of the large SUSY contributions, the mSUGRA predictions
for $B \to \pi^- \bar{K}^{*0}$, $\pi^0 K^{*-}$ and $\pi^0 K^{*0}$
become consistent with the data within one standard deviation.

\item 
For all the four $B \to\pi K^{*}$ decays, the dominant error is induced by the 
uncertainty of the annihilation contribution, it is large 
and may mask the large SUSY contributions in Case B.

\item 
After the inclusion of the large theoretical errors, the
theoretical predictions for the branching ratios of $B \to \pi^-
\bar{K}^{*0}$, $\pi^0 K^{*-}$ and $\pi^0 K^{*0}$ in both models
can be consistent with the corresponding data within one standard deviation. 
But for $\bar{B}^0 \to \pi^+ K^{*-}$ decay, only when we include the large SUSY
contribution in Case B as well can the theoretical value match
with the data  within one standard deviation.

\end{itemize}

\begin{figure}[tbp]
\centerline{\mbox{\epsfxsize=8cm\epsffile{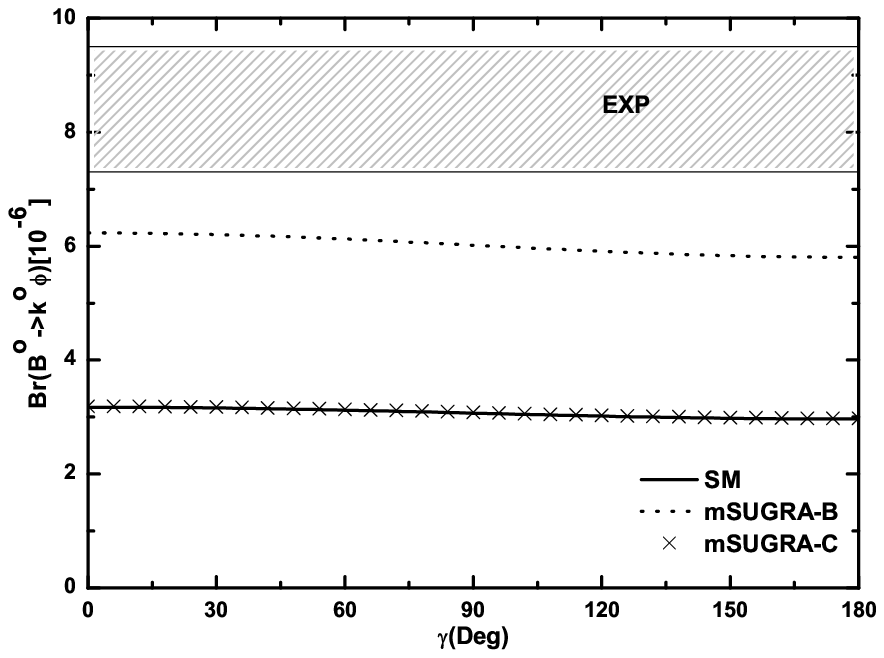}
\epsfxsize=8cm\epsffile{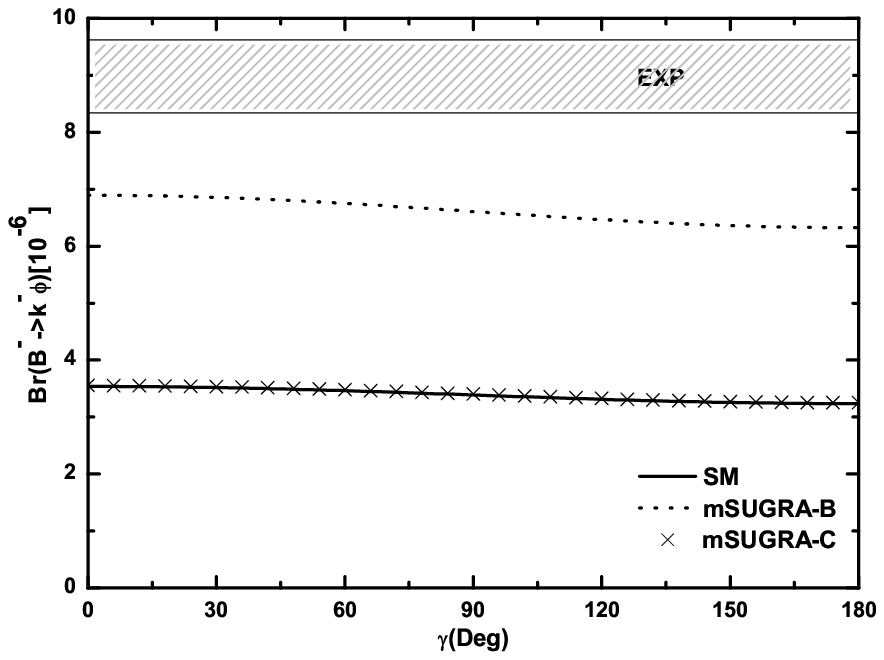}}}
\vspace{0.1cm}
\caption{The same as Fig.\ref{fig:pirho} but for $B \to K \phi$ decays.}
\label{fig:ko}
\end{figure}

Like $B \to \pi K^*$,  the SUSY contributions  to
$B\to K\phi$ in case B are large and  can provide about $100\%$ enhancement.
From Fig.~\ref{fig:ko}, one can see that the mSUGRA predictions in case B
can become consistent with the data within one standard
deviation. The data, the central values and the major errors of the
branching ratios (in units of $10^{-6}$) in the SM and mSUGRA model
are
\begin{eqnarray}
Br(\bar{B}^0 \to  \bar{K}^{0}\phi)&=& \left \{\begin{array}{ll}
 8.3 ^{+1.2}_{-1.0} , & {\rm  Data}, \\
 3.1 ^{+0.5}_{-0.6}  (\mu) \; ^{+1.1}_{-0.9} (F_1^{B\to K}) \;
 ^{+8.0}_{-2.2}(\chi_A),
 & {\rm SM}, \\
6.1\pm1.1 (\mu) \; ^{+2.1}_{-1.8} (F_1^{B\to K})\;
 ^{+10.2}_{-3.3}(\chi_A),
 & {\rm  Case-B}, \\
3.1 ^{+0.5}_{-0.6}  (\mu) \; ^{+1.1}_{-0.9} (F_1^{B\to K}) \;
 ^{+8.0}_{-2.2}(\chi_A),
 & {\rm  Case-C}, \\
\end{array} \right. \label{eq:ko1} \\
Br(B^- \to K^- \phi)&=& \left \{\begin{array}{ll}
 9.0 \pm 0.7 , & {\rm  Data}, \\
 3.5 ^{+0.6}_{-0.7} (\mu) \; ^{+1.2}_{-1.0} (F_1^{B\to K}) \;
 ^{+9.0}_{-2.4}(\chi_A),
 & {\rm  SM}, \\
6.8^{+1.3}_{-1.2}(\mu) \; ^{+2.3}_{-1.9} (F_1^{B\to K})
\;^{+11.5}_{-3.7}(\chi_A),
 & {\rm  Case-B}, \\
 3.5 ^{+0.6}_{-0.7} (\mu) \; ^{+1.2}_{-1.0} (F_1^{B\to K})\;
  ^{+9.0}_{-2.4}(\chi_A),
  & {\rm  Case-C}.\\
\end{array} \right. \label{eq:ko2}
\end{eqnarray}

From these numerical results, one can see that (a) the central values of the 
SM predictions are only about one third of the measured values. But the
mSUGRA predictions in case B can become consistent with the data
within one standard deviation. (b) Similar to the $B \to \pi K^*$ decays,
the dominant error here also comes from the uncertainty of 
the annihilation contribution and the form factor $F_1^{B \to K}$.
The error induced by $\chi_A$ are very large and even mask the
large SUSY contributions in Case B. 

\subsubsection{$B\to  K\rho$, $B\to K \omega$}

\begin{figure}[tbp]
\centerline{\mbox{\epsfxsize=18cm\epsffile{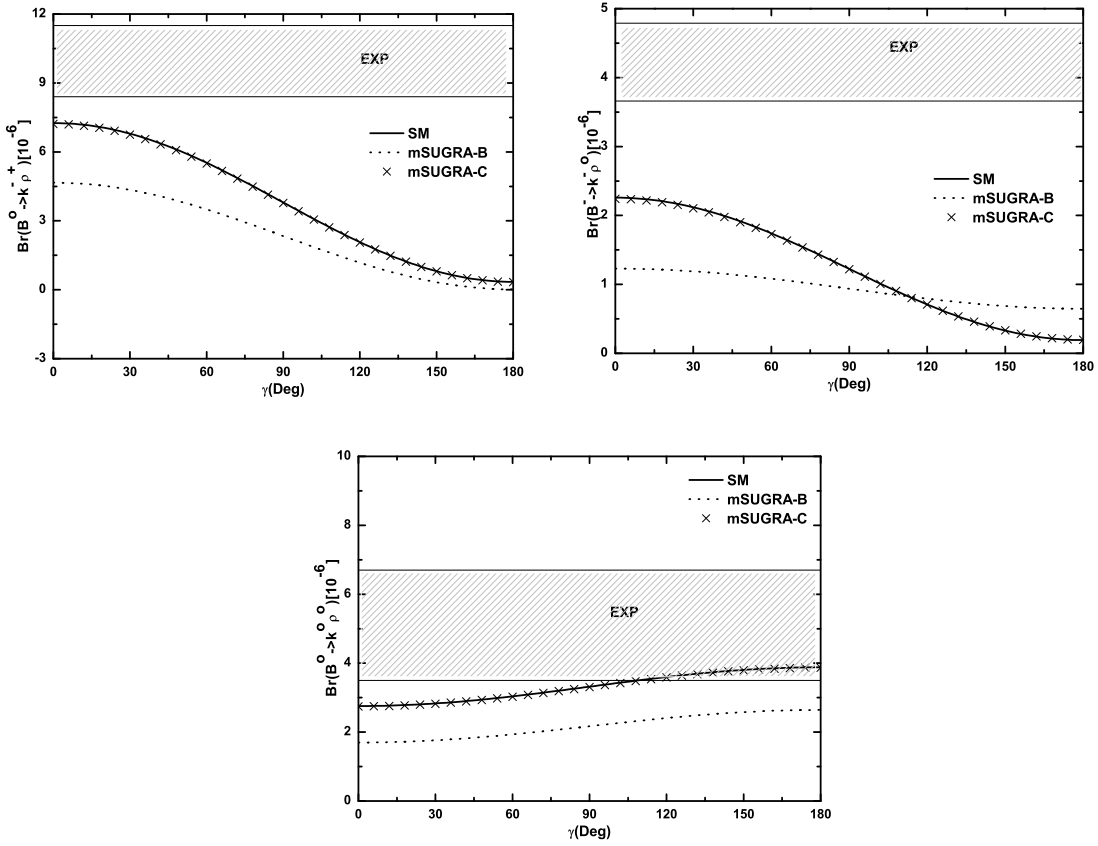}}}
\vspace{-1cm}
\caption{The same as Fig.\ref{fig:pirho} but for $B \to K \rho$ decays.}
\label{fig:krho}
\end{figure}

Among the four $B\to  K\rho$ decays, three
of them have been well measured experimentally.
The data, the theoretical predictions and the major errors
(in units of $10^{-6}$) are
\beq
Br(\bar{B}^0 \to  K^{-}\rho^+)&=& \left \{\begin{array}{ll}
 9.9 ^{+1.6}_{-1.5}, & {\rm  Data}, \\
 5.6 ^{+1.7}_{-1.5} (A_0^{B\to\rho})\;^{+0.9}_{-1.1}(\gamma)
  \;^{+3.8}_{-2.0} (\overline{m}_s) \;^{+8.1}_{-2.6}(\chi_A),
 & {\rm  SM}, \\
3.6^{+1.1}_{-0.9} (A_0^{B\to\rho})\;^{+0.6}_{-0.7}(\gamma)\;
 ^{+3.3}_{-1.5} (\overline{m}_s)\;^{+6.6}_{-1.9}(\chi_A),
 & {\rm Case-B}, \\
5.6 ^{+1.7}_{-1.5} (A_0^{B\to\rho})\;^{+0.9}_{-1.1}(\gamma)
  \;^{+3.8}_{-1.9} (\overline{m}_s)\;^{+8.1}_{-2.6}(\chi_A),
 & {\rm Case-C}, \\
\end{array} \right. \label{eq:krho1} \\
Br(B^- \to K^- \rho^0)&=& \left \{\begin{array}{ll}
 4.23 ^{+0.56}_{-0.57} , & {\rm  Data}, \\
  1.8 \pm 0.2 (F_1^{B\to K}) \;  ^{+0.7}_{-0.6} (A_0^{B\to\rho})
\;^{+1.5}_{-0.7} (\overline{m}_s)\;^{+3.1}_{-0.8}(\chi_A),
 & {\rm SM}, \\
1.1 \pm 0.1(F_1^{B\to K}) \; \pm0.4
(A_0^{B\to\rho})\;^{+1.2}_{-0.4}
(\overline{m}_s)\;^{+2.3}_{-0.4}(\chi_A),
 & {\rm Case-B}, \\
  1.8 \pm 0.2 (F_1^{B\to K}) \;  ^{+0.7}_{-0.6} (A_0^{B\to\rho})
\;^{+1.4}_{-0.7} (\overline{m}_s)\;^{+3.0}_{-0.8}(\chi_A),
 & {\rm Case-C},
\end{array} \right. \label{eq:krho2}\\
Br(\bar{B}^0 \to \bar{K}^0 \rho^0)&=& \left \{\begin{array}{ll}
 5.1 \pm 1.6 , & {\rm  Data}, \\
  3.0 \pm0.3(F_1^{B\to K}) \; \pm0.5 (A_0^{B\to\rho})
  \;^{+1.9}_{-1.1} (\overline{m}_s)\;^{+4.4}_{-1.5}(\chi_A),  & { \rm SM}, \\
1.9 \pm 0.3(F_1^{B\to K}) \; \pm0.2
(A_0^{B\to\rho})\;^{+1.7}_{-0.9}
(\overline{m}_s)\;^{+3.7}_{-1.1}(\chi_A),
 & { \rm Case-B}, \\
 3.0 \pm0.3(F_1^{B\to K}) \; \pm0.5 (A_0^{B\to\rho})
  \;^{+1.9}_{-1.0} (\overline{m}_s)\;^{+4.3}_{-1.5}(\chi_A),
 & {\rm Case-C}.\\
\end{array} \right. \label{eq:krho3}
\eeq
Here the new dominant error source is the uncertainty of the mass $\overline{m}_s$,
$85$ MeV $\leq \overline{m}_s \leq 125$ MeV.
In Fig.~\ref{fig:krho}, we show  the $\gamma$ dependence of the branching
ratios for the three measured $B\to  K\rho$ decays respectively.

From Eqs.(\ref{eq:krho1}-\ref{eq:krho3}) and Fig.~\ref{fig:krho},
one can see that the central values of the theoretical predictions in the SM and the
mSUGRA model of case C are almost identical and about half of the
measured values. In case B, unfortunately, the SUSY contribution
produce a further forty percent variation in the ``wrong"
direction. But when we include the large errors induced by the 
uncertainties of $\chi_A$, $\overline{m}_s$ and form factors, 
the theoretical predictions in both the SM and the
mSUGRA model (both Case B and Case C) can be consistent with the
data.

\begin{figure}[tbp]
\centerline{\mbox{\epsfxsize=8cm\epsffile{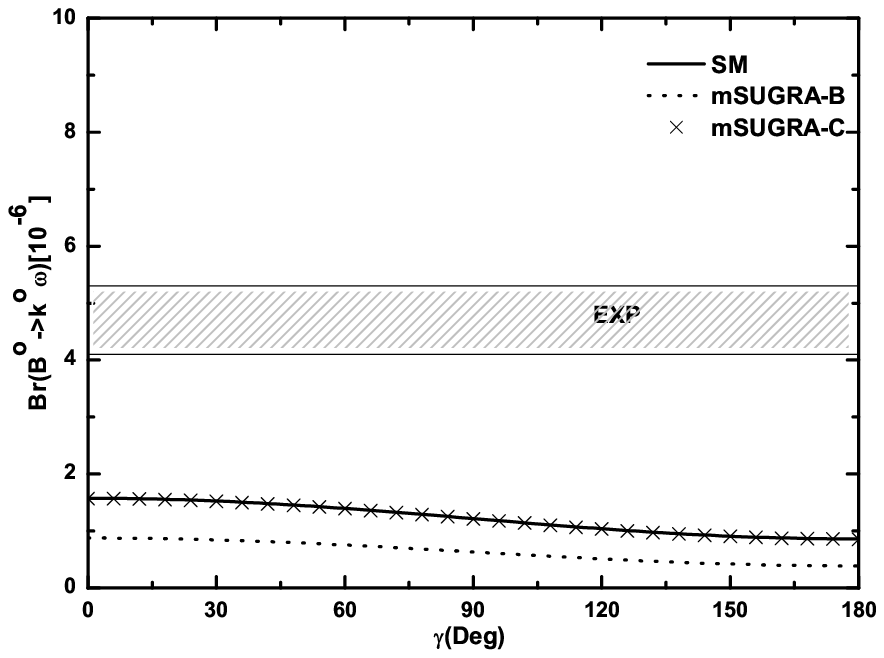}
\epsfxsize=8cm\epsffile{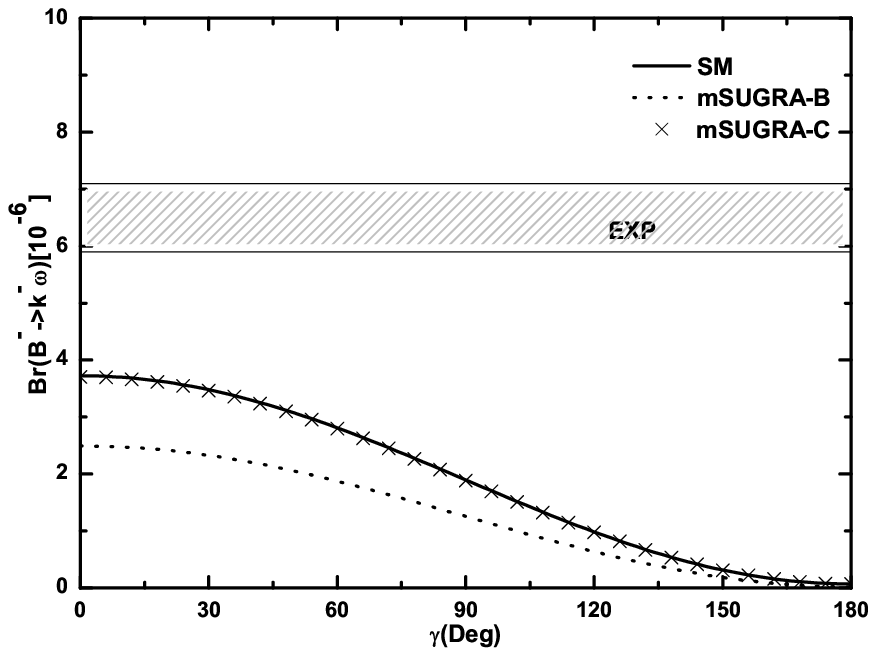}}}
\vspace{0.2cm}
\caption{The same as Fig.\ref{fig:pirho} but for $B \to K \omega$ decays.}
\label{fig:kw}
\end{figure}

For $B\to K \omega$ decays, the situation is very similar to the
$B \to K\rho$ decays. From Eqs.(\ref{eq:kw1}-\ref{eq:kw2}) and
Fig.~\ref{fig:kw}, one can see that the central values of
theoretical predictions in both the SM and the mSUGRA model are
much smaller than the measured values. And the SUSY corrections in
Case B even lead to a further reduction of the branching ratios.
The data and the theoretical predictions (in units of $10^{-6}$)
are
\begin{eqnarray}
&Br(\bar{B}^0& \to  \bar{K}^{0}\omega)= \left \{\begin{array}{ll}
 4.7 \pm 0.6 , & {\rm  Data}, \\
 1.4 \pm0.3(A_0^{B\to\omega}) \;^{+1.3}_{-0.7} (\overline{m}_s)\;
^{+3.0}_{-0.8}(\chi_A),
 & {\rm  SM}, \\
0.8\pm0.1(A_0^{B\to\omega})  \;^{+1.1}_{-0.5}
(\overline{m}_s)\;^{+2.4}_{-0.6}(\chi_A),
 & {\rm  Case-B}, \\
 1.4\pm0.3(A_0^{B\to\omega}) \;^{+1.3}_{-0.6} (\overline{m}_s)\;
 ^{+3.0}_{-0.8}(\chi_A),
 & {\rm Case-C}, \\
\end{array} \right. \label{eq:kw1} \\
&Br(B^-& \to K^- \omega)= \left \{\begin{array}{ll}
 6.5 \pm 0.6 , & {\rm  Data}, \\
 2.9 \pm0.7(A_0^{B\to\omega})\;^{+0.5}_{-0.6}(\gamma)
  \;^{+1.8}_{-1.0} (\overline{m}_s)\;^{+3.7}_{-1.2}(\chi_A),
 & {\rm SM}, \\
 1.9^{+0.5}_{-0.4}(A_0^{B\to\omega}) \;^{+0.3}_{-0.4}(\gamma)
 \;^{+1.6}_{-0.8} (\overline{m}_s)\;^{+3.0}_{-0.9}(\chi_A),
 & {\rm  Case-B}, \\
  2.9 \pm0.7(A_0^{B\to\omega})\;^{+0.4}_{-0.6}(\gamma)
  \;^{+1.8}_{-1.0} (\overline{m}_s)\;^{+3.7}_{-1.2}(\chi_A),
 & {\rm Case-C}. \\
\end{array} \right. \label{eq:kw2}
\end{eqnarray}
Here one can see that (a) the central values of the theoretical predictions in 
both the SM and mSUGRA model are all smaller than the data; 
and (b) the dominant error still comes from the uncertainty of the 
annihilation contributions and is very large in size.

\subsubsection{$B\to K^* \eta$, $B\to  \eta^{'}\rho$ }

\begin{figure}[tbp]
\vspace{-1cm}
\centerline{\mbox{\epsfxsize=18cm\epsffile{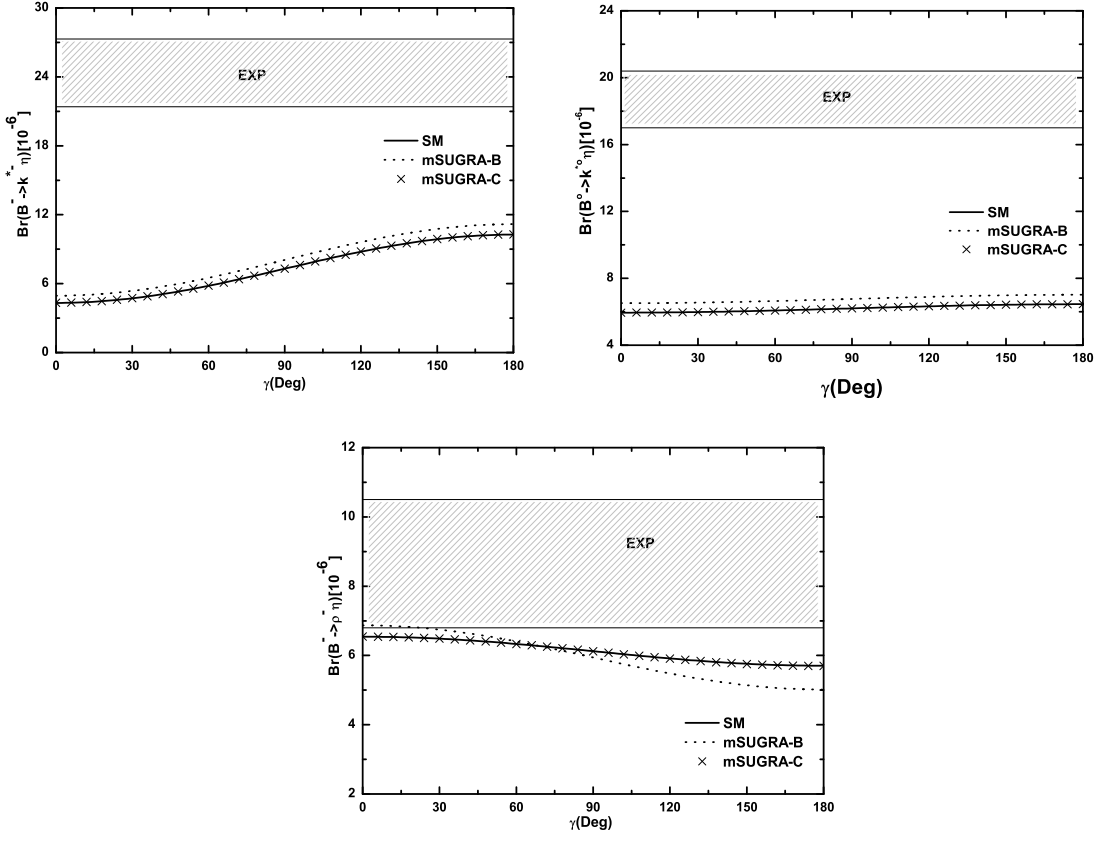}}}
\vspace{-1cm}
\caption{The same as Fig.\ref{fig:pirho} but for $B \to K^*\eta$ and
$B^-\to \rho^- \eta$ decays.}
\label{fig:etap}
\end{figure}

For the decay channels involving $\eta^{(')}$ meson,
the dynamics is rather complex and  has been studied
by many authors, for example, in Refs.~\cite{bn03a,ajcp98}. Here
we didn't consider the additional form-factor type contribution to
the flavor-singlet coefficients $\alpha_3^p(M_1\eta^{(')}_{q,s})$
( see Ref.~\cite{bn03b} ) since it has large uncertainties. The
data and the theoretical predictions (in units of $10^{-6}$)
for the three measured decay modes in both the SM and the mSUGRA model are
\beq
Br(B^- \to  K^{*-}\eta)&=& \left \{\begin{array}{ll}
 24.3 ^{+3.0}_{-2.9} , & {\rm  Data}, \\
 5.7 ^{+1.0}_{-0.9} (A_0^{B\to K ^{*}})
\;^{+1.0}_{-0.8} (\gamma) \;  ^{+3.7}_{-2.0}
(\overline{m}_s)\;^{+9.0}_{-2.9}(\chi_A),
 & {\rm  SM}, \\
6.4 \pm0.8(A_0^{B\to K ^{*}})\;
  \;^{+1.0}_{-0.8}(\gamma) ^{+4.3}_{-2.3}(\overline{m}_s)\;
  ^{+9.4}_{-3.1}(\chi_A),
 & {\rm Case-B}, \\
 5.7 ^{+1.0}_{-0.9} (A_0^{B\to K ^{*}})
\;^{+1.0}_{-0.8} (\gamma) \;  ^{+3.7}_{-2.0}
(\overline{m}_s)\;^{+9.0}_{-2.9}(\chi_A),
 & {\rm Case-C}, \\
\end{array} \right. \label{eq:etakp1}
\eeq
\beq
Br(\bar{B}^0 \to  \bar{K}^{*0}\eta)&=& \left \{\begin{array}{ll}
18.7 \pm 1.7 , & {\rm  Data}, \\
 6.1 ^{+1.0}_{-0.9} (A_0^{B\to K ^{*}})\;
\;\pm0.1 (\gamma) ^{+3.7}_{-2.0}
(\overline{m}_s)\;^{+9.1}_{-3.0}(\chi_A),
 & {\rm SM}, \\
6.6 \pm0.8(A_0^{B\to K ^{*}})\;
 \;\pm0.1(\gamma)^{+4.2}_{-2.3}(\overline{m}_s)\;^{+9.5}_{-3.2}(\chi_A),
 & {\rm Case-B}, \\
 6.1 ^{+1.0}_{-0.9} (A_0^{B\to K ^{*}})\;
\;\pm0.1 (\gamma) ^{+3.7}_{-2.0}
(\overline{m}_s)\;^{+9.1}_{-3.0}(\chi_A),
 & {\rm Case-C}, \\
\end{array} \right. \label{eq:etakp2}\\
Br(B^- \to  \rho^-\eta)&=& \left \{\begin{array}{ll}
 8.6 ^{+1.9}_{-1.8} , & {\rm  Data}, \\
 6.3 \pm0.2(\mu) \;\pm 0.1 (\gamma)
 \; \pm0.1 (\overline{m}_s)\;\pm0.3(\chi_A),
 & {\rm SM}, \\
6.4 ^{+0.2}_{-0.1}  (\mu) \;
  ^{+0.2}_{-0.3} (\gamma)\;\pm0.1(\overline{m}_s)\;\pm0.3(\chi_A),
 & {\rm Case-B}, \\
6.3 \pm0.2(\mu) \;\pm 0.1 (\gamma)
 \; \pm0.1 (\overline{m}_s)\;\pm0.3(\chi_A),
 & {\rm Case-C}. \\
\end{array} \right. \label{eq:etarho1}
\eeq
Clearly the uncertainties of $\chi_A$  and $\overline{m}_s$
are the dominant error sources here.

By comparing the numerical results with the measured values, one finds that
\begin{itemize}

\item 
The SUSY contributions to these three measured decays are
always small for both case B and C. For $B^-\to K^{*-} \eta$ and
$\overline{B}^0 \to \overline{K}^{*0} \eta$ decays, the
central values of the theoretical predictions in both models are
only around $30\%$ of the measured values, and the data can be
explained only when we take the large theoretical and experimental 
errors into account. For $B^- \to K^- \eta$ and $\overline{B}^0
\to \overline{K}^0 \eta$ decays, however, the theoretical
predictions in the SM and mSUGRA model as given in Ref.~\cite{zw04} 
agree well with the data.

\item 
For $B^-\to\rho^-\eta$ decay, a $b\to d$ transition process,
the SUSY contribution is very  small, and the theoretical predictions in the 
SM and mSUGRA model are consistent with the data within one standard 
deviation.

\item 
As illustrated in Fig.~\ref{fig:etap}, the branching ratios
of all the three channels have a weak dependence on the angle
$\gamma$. For $B\to K^* \eta$ decays, the theoretical errors are large in 
size and induced dominantly by the uncertainty of the parameter $\chi_A$ and the mass
$\overline{m}_s$. 
For $B^-\to \rho^-\eta$ decay, however, the theoretical errors are relatively 
small.

\end{itemize}

\section{Branching ratios for $B\to VV$ decays}\label{sec:vv}

For $B\to VV$ decay modes, one generally should evaluate three
amplitudes with different helicity since they all can make
contributions and do not interfere with each other. In terms of the
helicity matrix elements
\beq
H_\lambda = <V_1(\lambda)V_2(\lambda)|H_{eff}|B)>, \ \ \lambda=0, \pm 1,
\eeq
the branching ratios of  $B\to VV$ decays can be written as
\beq
Br(B \to V_1 V_2 ) =  \tau_{B}\, \frac{|p_c|}{8\pi
M_B^2}\left [ |H_0|^2 + |H_{+1}|^2 + |H_{-1}|^2 \right ] ~.
\eeq
The three independent helicity amplitudes $H_0$, $H_{+1}$ and
$H_{-1}$ can be expressed by three invariant amplitudes
$a_\lambda, b_\lambda, c_\lambda$ defined by the decomposition
\beq
H_\lambda = i\epsilon^\mu(\lambda)\eta^\nu(\lambda)\left[
    a_\lambda g_{\mu\nu}+\frac{b_\lambda}{M_1 M_2}p_\mu
    p_\nu + \frac{ic_\lambda}{M_1 M_2}\epsilon_{\mu\nu\alpha\beta}p_1^\alpha
    p_2^\beta \right]. \label{eq:hl}
\eeq Here  $\epsilon^\mu(\eta^\nu)$, $p_{1,2}$ and $M_{1,2}$ are
the polarization vector, four momentum and masses of $V_{1,2}$,
respectively, while $p=p_1 + p_2 $ is the four-momentum of B
meson. The helicity elements $H_\lambda$ can be further simplified
as \beq H_{\pm1} &=& a_{\pm1} \pm c_{\pm1} \sqrt{x^2-1}, ~~~~
H_0 = -a_0x - b_0\left ( x^2-1 \right ) \label{eq:h01} \nonumber\\
x &=& \frac{M_B^2-M_1^2-M_2^2}{2M_1 M_2}
\eeq

\subsection{Helicity amplitude of $B \to \rho^+ \rho^-$, an example}

Now we take  the decay $\bar{B}^0\to \rho^{\pm}\rho^{\mp}$ as an
example to show the ways of decomposition. With the QCD factorization approach,
the decay amplitude of $\bar{B}^0\to \rho^{\pm}\rho^{\mp}$ decay
reads
\beq
{\cal  A}^\lambda (B \to \rho^+\rho^-)&=&-i \frac{G_F}{ \sqrt 2 }f_{\rho}m_{\rho}
\left [
 (\varepsilon _ +   \cdot \varepsilon _ - )(m_B  + m_\rho )A_1^{B
\to \rho } (m_\rho ^2 ) \right. \non
&& \left. - (\varepsilon_ + \cdot p_B )(\varepsilon
_ -   \cdot p_B )\frac{2A_2^{B \to \rho } (m_\rho ^2 )}{m_B  +
m_\rho  }
-i  \varepsilon _{\mu \nu
\alpha \beta } \varepsilon _ - ^\mu \varepsilon _ + ^\nu
p_B^\alpha  p_ + ^\beta  \frac{V^{B \to \rho } (m_\rho ^2
)}{{m_B  + m_\rho  }}\right ] \non
&&\cdot  \left [ V_{ub} V_{ud}^* a_1^\lambda   -
V_{tb} V_{td}^* \left (a_4^\lambda + a_{10}^\lambda \right ) \right]
\eeq
By comparing this expression of decay amplitude ${\cal A}_\lambda$ with
Eq.(\ref{eq:hl}), one can  find the coefficients $a_\lambda, b_\lambda$
and $c_\lambda$,
\beq
a_\lambda(B \to \rho^+ \rho^-)&=&-\frac{G_F}{ \sqrt 2 }f_{\rho}m_{\rho}(m_B  + m_\rho
)A_1^{B \to \rho } (m_\rho ^2 )\{ V_{ub} V_{ud}^* a_1^\lambda   -
V_{tb} V_{td}^* [a_4^\lambda + a_{10}^\lambda  ]\},   \non
b_\lambda(B \to \rho^+ \rho^-)&=&{\sqrt 2} G_F f_{\rho}m_{\rho}^3 \frac{{A_2^{B \to
\rho } (m_\rho ^2 )}}{{m_B  + m_\rho  }} \{ V_{ub} V_{ud}^*
a_1^\lambda   - V_{tb} V_{td}^* [a_4^\lambda + a_{10}^\lambda  ]\},
 \non
c_\lambda(B \to \rho^+ \rho^-)&=&{\sqrt 2} G_F f_{\rho}m_{\rho}^3 \frac{{V^{B \to \rho
} (m_\rho ^2 )}}{{m_B  + m_\rho  }} \{ V_{ub} V_{ud}^* a_1^\lambda
- V_{tb} V_{td}^* [a_4^\lambda + a_{10}^\lambda  ]\},
\end{eqnarray}
where the factorized coefficients $a_1^\lambda$, $a_4^\lambda$ and
$a_{10}^\lambda$ can be written as
\begin{eqnarray}\label{a1l}
a_1^\lambda   &=& C_1  + \frac{{C_2 }}{{N_c }} + \frac{{\alpha _s
}}{{4\pi }}\frac{{C_F }}{{N_c }}C_2 \left[ {f_I^\lambda   +
f_{II}^\lambda  } \right], \\
a_4^\lambda   &=& C_4  + \frac{{C_3 }}{{N_c }} + \frac{{\alpha _s
}}{{4\pi }}\frac{{C_F }}{{N_c }}C_3 \left[ {f_I^\lambda  +
f_{II}^\lambda  } \right] + \frac{{\alpha _s }}{{4\pi }}\frac{{C_F
}}{{N_c }}\left\{ - C_1 \left[ {\frac{{\upsilon _u }}{{\upsilon _t
}}G^\lambda  (s_u ) + \frac{{\upsilon _c
}}{{\upsilon _t }}G^\lambda  (s_c )} \right] \right. \nonumber \\
&\ \ &+C_3 \left[ {G^\lambda  (s_q ) + G^\lambda  (s_b )} \right]
+ (C_4  + C_6 )\sum\limits_{q^{'}=u}^b {\left[ {G^\lambda
(s_{q^{'} } - \frac{2}{3})} \right]}  + \frac{3}{2}C_9 \left[
e_qG^\lambda  (s_q ) + \right. \nonumber \\
&\ \ &\left.\left.e_b G^\lambda  (s_b ) \right]+ \frac{3}{2}(C_8 +
C_{10} )\sum\limits_{q^{'}  = u}^b {e_{q^{'} } \left[ {G^\lambda
(s_{q^{'} }  - \frac{2}{3})}
\right]}+C_{8g} G_g^\lambda\right\},\\
a_{10}^\lambda  &=& C_{10}  + \frac{{C_9 }}{{N_c }} +
\frac{{\alpha _s }}{{4\pi }}\frac{{C_F }}{{N_c }}C_9 \left[
{f_I^\lambda  + f_{II}^\lambda } \right] - \frac{{\alpha _e
}}{{9\pi }}C_e^\lambda,\label{a2l}
\end{eqnarray}

In the above equations, the vertex corrections $f_I^{\lambda}$ and
the hard spectator scattering contributions $f_{II}^\lambda$ are
given by
 \begin{eqnarray}
 f_I^0&=&-12\ln \frac{\mu}{m_b}-18+\int_0^1{\rm d}u
\Phi_{\parallel}^{V_2} (u)\left(3\frac{1-2u}{1-u}\ln u-3i \pi\right),\\
 f_I^{\pm}&=&-12\ln \frac{\mu}{m_b}-18+\int_0^1{\rm d}u
\left(g_{\perp}^{(v){V_2}} (u)\pm \frac{a
g_{\perp}^{\prime(a){V_2}}
(u)}{4}\right)\left(3\frac{1-2u}{1-u}\ln u-3i
\pi\right),\\
f^{0}_{II}&=&\frac{4\pi^2}{N_C}\frac{if_{B}
f_{V_1}f_{V_2}}{h_0}\int_0^1{\rm d}\xi \frac{\Phi_1^B
(\xi)}{\xi}\int_0^1{\rm d}v \frac{\Phi_{\parallel}^{V_1}
(v)}{\bar{v}}\int_0^1{\rm d}u \frac{\Phi_{\parallel}^{V_2}
(u)}{u},\\
f_{II}^{\pm}&=&-\frac{4
\pi^2}{N_C}\frac{2if_{B}f^{\perp}_{V_1}f_{V_2}m_{V_2}}{m_{B}h_{\pm}}
(1\mp1)\int_0^1{\rm d}\xi \frac{\Phi_1^B (\xi)}{\xi}\int_0^1{\rm
d}v \frac{\Phi_{\perp}^{V_1} (v)}{\bar{v}^2}\nonumber
\\&&\times\int_0^1{\rm d}u \left(g_{\perp}^{(v){V_2}}
(u)-\frac{g_{\perp}^{\prime(a){V_2}} (u)}{4}\right)+\frac{4
\pi^2}{N_C}\frac{2if_{B}f_{V_1}f_{V_2}m_{V_1}m_{V_2}}{m_{B}^2h_{\pm}}
\int_0^1{\rm d}\xi \frac{\Phi_1^B (\xi)}{\xi}\nonumber
\\&&\times\int_0^1{\rm d}v {\rm d}u\left(g_{\perp}^{(v){V_1}} (v)
\pm\frac{g_{\perp}^{\prime(a){V_1}}
(v)}{4}\right)\left(g_{\perp}^{(v){V_2}} (u)\pm \frac{
g_{\perp}^{\prime(a){V_2}} (u)}{4}\right)\frac{u+\bar v} {u{\bar
v}^2}, \label{hard}
\end{eqnarray}
here $\bar{v}=1-v$ and the light-cone distribution amplitudes
(LCDAS) $\Phi_{\parallel}^V(u)$, $\Phi_{\perp}^V(u)$,
$g_{\perp}^{(a)V}$,  $g_{\perp}^{(v)V}$ can be found in
Ref.~\cite{vvqcd}.

The functions describing the contributions of the QCD penguin-type
diagrams,  the dipole operator $O_{8g}$  and the electro-weak
penguin-type diagrams in eqs.(\ref{a1l}-\ref{a2l}) are
\cite{vvqcd}
\beq G^0 (s)&=&\frac{2}{3}-\frac{4}{3}\ln
\frac{\mu}{m_b}+4\int_0^1{\rm d}u ~\Phi_{\parallel}^{V_2}
(u) g(u,s),\\
G^{\pm}(s)&=&\frac{2}{3}-\frac{2}{3}\ln
\frac{\mu}{m_b}+2\int_0^1{\rm d}u ~(g_{\perp}^{(v){V_2}}(u)\pm
\frac{g_{\perp}^{\prime(a){V_2}}(u)}{4}) g(u,s),
\eeq
\beq
G_g^0&=&-\int_0^1 {\rm d}u ~\frac{2\Phi_{\parallel}^{V_2}
(u)}{1-u},\\
G_g^+&=&-\int_0^1{\rm d}u ~\left( g_{\perp}^{(v){V_2}}(u)+
\frac{g_{\perp}^{\prime(a){V_2}}(u)}{4}\right)\frac{1}{1-u},\\
G_g^-&=&\int_0^1 \frac{{\rm d}u}{\bar{u}}\left[
-\bar{u}g_{\perp}^{(v){V_2}}(u)+
\frac{\bar{u}g_{\perp}^{\prime(a){V_2}}(u)}{4} \right.\non
&& \left. +\int_0^u{\rm
d}v\left(\Phi_\parallel^{V_2}(v)-g_\perp^{(v)V_2}(v)\right)
+\frac{g_{\perp}^{(a){V_2}}(u)}{4}\right]\label{o8g},\\
 C^\lambda_e&=&\left[\frac{v_u}{v_t}G^\lambda (s_u)+\frac{v_c}{v_t}
 G^\lambda (s_c)\right]\left(
C_2+\frac{C_1}{N_C}\right),
 \eeq
with the function $g(u,s)$ defined as
 \begin{eqnarray}
 g(u,s)&=&\int_0^1{\rm d}x~ x\bar{x}\ln{(s-x\bar{x}\bar{u} -i \epsilon)}.
 \end{eqnarray}

From Eq.(\ref{eq:h01}), we can finally get the helicity elements
$H_{\pm 1}$ and $H_{0}$. One should note that the coefficients
$a,b$ and $c$ are independent of the helicity $\lambda$ in the
naive factorization approach. In QCD factorization approach,
however, the coefficients $a_\lambda, b_\lambda$ and $c_\lambda$
depend on the choice of $\lambda$ ($\lambda=0,\pm 1$).

\subsection{Numerical results}

Using the above formulae it is straightforward to calculate the
branching ratios of $B\to VV$ decays. In Table
\ref{smmssm3} and \ref{smmssm4}, we show the theoretical predictions
for the CP-averaged branching ratios of the nineteen  $B\to VV$
decays in both the SM and the mSUGRA model, assuming $\mu=m_b/2,
m_b$ and $2 m_b$, respectively. 

In numerical calculations, we do not consider the annihilation contributions 
and only give $Br^{f}$. The reasons for this can be found in the last paragraph 
of Sec.~\ref{sec:qcdf}. Of course, the annihilation contribution may 
play an important role to explain the large transverse polarization of 
the $B \to K^{*0} \rho^+$ and $\phi K^{*+}$ decays, as pointed out in a 
a recent paper \cite{kagan04}, where the author has tried to include the contribution 
induced by a QCD penguin annihilation graph to explain the observed longitudinal 
polarization, $f_L(B \to \phi K^*) \approx 50\%$, in the framework of the SM. 
More studies are needed to understand this problem clearly. 
But it is beyond the scope of this paper, and we here just calculate the possible 
new physics contributions 
to the CP-averaged branching ratios of $B \to VV$ decays in the mSUGRA model, and 
show the general pattern of such contributions.

From the numerical results as given in Table \ref{smmssm3} and \ref{smmssm4}, 
one can see that the new physics contributions to the branching ratios of $B \to VV$
decays are generally small for the SM-like case-C, but
can be rather large for the case-B, especially to those penguin dominated
$B\to K^*(\rho, \omega, \phi)$ ( $b\to s$ transition ) processes.

Among the nineteen $B \to VV$ decays, only seven of
them have been well measured, while upper limits are available for
the remaining twelve decay modes. For the world average of the
experimental measurements, one can see HFAG home page \cite{hfag} and
references therein.
In the following subsections we focus on the seven measured decay channels.

\begin{table}[tbp]
\doublerulesep 1.5pt
\caption{Numerical predictions in the  SM and
mSUGRA model (both case B and case C) for
CP-averaged branching ratios in units of $10^{-6}$ for $b\to d$
transition processes of $B\to VV$ decays, here the annihilation
contributions are not included. } \label{smmssm3}
\begin{center}
\begin{tabular} {l|c|cc|c|cc|c|cc} \hline  \hline
\multicolumn{1}{c|}{$B\to VV$} &
\multicolumn{3}{|c|}{$\mu=m_{b}/2$} &
\multicolumn{3}{|c|}{$\mu=m_{b}$}&
\multicolumn{3}{|c}{$\mu=2m_{b}$}
 \\ \cline{2-10}
\ \ \ \ $(b\to d)$& \multicolumn{1}{|c|}{\ \ \ \ SM \ \ \ \ }&
\multicolumn{2}{|c|}{mSUGRA} &
       \multicolumn{1}{|c|}{\ \ \ \ SM \ \ \ \ }& \multicolumn{2}{|c|}{mSUGRA} &
      \multicolumn{1}{|c|}{\ \ \ \ SM \ \ \ \ }& \multicolumn{2}{|c}{mSUGRA}
  \\ \cline{2-10}
\   \ & \multicolumn{1}{|c|}{\ \ \ \ $Br^{f}$ \ \ \ \ }& (B)&(C) &
       \multicolumn{1}{|c|}{\ \ \ \ $Br^{f}$ \ \ \ \ }& (B)&(C) &
      \multicolumn{1}{|c|}{\ \ \ \ $Br^{f}$ \ \ \ \ }& (B)&(C)
  \\ \hline\hline
$\bar{B}^0\to\rho^+\rho^-$&27.8&28.1&27.8&27.5&27.8&27.5&27.1&27.3&27.1\\
$B^-\to\rho^-\rho^0$&18.4&18.4&18.4&18.7&18.7&18.7&19.1&19.1&19.1\\
$B^-\to\rho^-\omega$&16.6&17.0&16.6&16.6&16.9&16.6&16.8&17.1&16.8\\
\hline \hline
$\bar{B}^0\to\rho^0\rho^0$&0.38&0.40&0.38&0.33&0.34&0.33&0.35&0.36&0.35\\
$\bar{B}^0\to\rho^0\omega$&0.09&0.17&0.09&0.07&0.13&0.07&0.05&0.11&0.05\\
$\bar{B}^0\to\omega\omega$&0.40&0.44&0.40&0.33&0.37&0.33&0.33&0.37&0.34\\
\hline \hline
$\bar{B}^0\to\rho^0\phi$&0.004&0.004&0.004&0.003&0.003&0.003&0.003&0.003&0.003\\
$B^-\to\rho^-\phi$&0.009&0.009&0.009&0.007&0.007&0.007&0.006&0.006&0.006\\
$\bar{B}^0\to\omega\phi$&0.003&0.003&0.003&0.003&0.003&0.003&0.002&0.002&0.002\\
\hline \hline
$\bar{B}^0\to \bar{K}^{*0}K^{*0}$&0.28&0.46&0.28&0.22&0.37&0.22&0.17&0.30&0.18\\
$B^-\to K^{*-}K^{*0}$&0.30&0.50&0.30&0.24&0.40&0.24&0.19&0.33&0.19\\
\hline \hline
\end{tabular}
\end{center}
\end{table}

\begin{table}[tbp]
\doublerulesep 1.5pt
\caption{Same as Table \ref{smmssm3}, but for $b\to s$
transition processes of $B\to VV$ decays.} \label{smmssm4}
\begin{center}
\begin{tabular} {l|c|cc|c|cc|c|cc} \hline  \hline
\multicolumn{1}{c|}{$B\to VV$} &
\multicolumn{3}{|c|}{$\mu=m_{b}/2$} &
\multicolumn{3}{|c|}{$\mu=m_{b}$}&
\multicolumn{3}{|c}{$\mu=2m_{b}$}
 \\ \cline{2-10}
\ \ \ \ $(b\to s)$& \multicolumn{1}{|c|}{\ \ \ \ SM \ \ \ \ }&
\multicolumn{2}{|c|}{mSUGRA} &
       \multicolumn{1}{|c|}{\ \ \ \ SM \ \ \ \ }& \multicolumn{2}{|c|}{mSUGRA} &
      \multicolumn{1}{|c|}{\ \ \ \ SM \ \ \ \ }& \multicolumn{2}{|c}{mSUGRA}
  \\ \cline{2-10}
\   \ & \multicolumn{1}{|c|}{\ \ \ \ $Br^{f}$ \ \ \ \ }& (B)&(C) &
       \multicolumn{1}{|c|}{\ \ \ \ $Br^{f}$ \ \ \ \ }& (B)&(C) &
      \multicolumn{1}{|c|}{\ \ \ \ $Br^{f}$ \ \ \ \ }& (B)&(C)
  \\ \hline\hline
$\bar{B}^0\to K^{*-}\rho^+$&3.74&6.44&3.76&3.11&5.32&3.13&2.60&4.44&2.61\\
$\bar{B}^0\to \bar{K}^{*0}\rho^0$&0.81&1.90&0.82&0.57&1.42&0.57&0.38&1.04&0.38\\
$B^-\to K^{*-}\rho^0$&4.43&6.63&4.44&3.87&5.74&3.88&3.40&5.02&3.41\\
$B^-\to K^{*0}\rho^-$&5.38&9.15&5.41&4.36&7.51&4.38&3.48&6.14&3.50\\
\hline
$\bar{B}^0\to K^{*0}\omega$&2.35&3.76&2.36&1.90&3.12&1.90&1.50&2.58&1.51\\
$B^-\to K^{*-}\omega$&2.02&3.19&2.03&1.70&2.71&1.71&1.45&2.32&1.45\\
\hline
$\bar{B}^0\to K^{*0}\phi$&5.61&9.91&5.64&4.24&7.82&4.26&3.13&6.15&3.15\\
$B^-\to K^{*-}\phi$&6.09&10.76&6.12&4.60&8.50&4.63&3.40&6.68&3.42\\
\hline \hline
\end{tabular}
\end{center}
\end{table}

\subsubsection{$\bar{B}^0\to \rho^{+}\rho^{-}$, $B^-\to \rho^{-}(\rho^{0}, \omega)$}

These three decays are  tree-dominated decay channels. The data and the theoretical
predictions of the branching ratios (in units of
$10^{-6}$) in the SM and  mSUGRA model are
\begin{eqnarray}
Br(\bar{B}^0 \to  \rho^{+}\rho^{-})&=& \left \{\begin{array}{ll}
 26.2 ^{+3.6}_{-3.7} , & {\rm  Data}, \\
 27.5 ^{+4.1}_{-3.8} (A^{B\to \rho})\;
^{+1.1}_{-1.4} (\gamma), & {\rm SM}, \\
27.8 ^{+4.1}_{-3.8} (A^{B\to \rho})\;
 \;^{+1.4}_{-1.8}(\gamma),
  & {\rm Case-B}, \\
 27.5 ^{+4.1}_{-3.8} (A^{B\to \rho})\;
^{+1.1}_{-1.4} (\gamma), & {\rm Case-C}, \\
\end{array} \right. \label{eq:rhorho1}\\
Br(B^- \to  \rho^-\rho^{0})&=& \left \{\begin{array}{ll}
 26.4 ^{+6.1}_{-6.4} , & {\rm  Data}, \\
18.7^{+2.5}_{-2.3} (A^{B\to \rho})\;
 ^{+0.3}_{-0.4} (\gamma), & {\rm  SM}, \\
 18.7^{+2.5}_{-2.3} (A^{B\to \rho})\;
 ^{+0.3}_{-0.4} (\gamma),
 & {\rm Case-B}, \\
18.7^{+2.5}_{-2.3} (A^{B\to \rho})\;
 ^{+0.3}_{-0.4} (\gamma),
 & {\rm Case-C}, \\
\end{array} \right. \label{eq:rhorho2}\\
Br(B^ -   \to \rho ^ -  \omega ) &=& \left\{ {\begin{array}{ll}
12.6 ^{+4.0}_{-3.7} , & {\rm  Data}, \\
16.6 _{ -2.1}^{ + 2.3}
(A^{B \to \omega } )_{ - 1.4}^{ + 1.0} (\gamma ), & {\rm SM}, \\
16.9  _{ - 2.2}^{ + 2.3} (A^{B \to \omega } )_{ - 1.8}^{ + 1.4}
(\gamma ),  & {\rm Case-B}, \\
16.6 _{ -2.1}^{ + 2.3}
(A^{B \to \omega } )_{ - 1.4}^{ + 1.1} (\gamma ), & {\rm Case-C}. \\
\end{array}} \right.
 \label{eq:rhow}
\end{eqnarray}
Here the dominant errors also come from the uncertainties of
the form factors and the angle $\gamma$.

\begin{figure}[tbp]
\vspace{-1cm}
\centerline{\mbox{\epsfxsize=20cm\epsffile{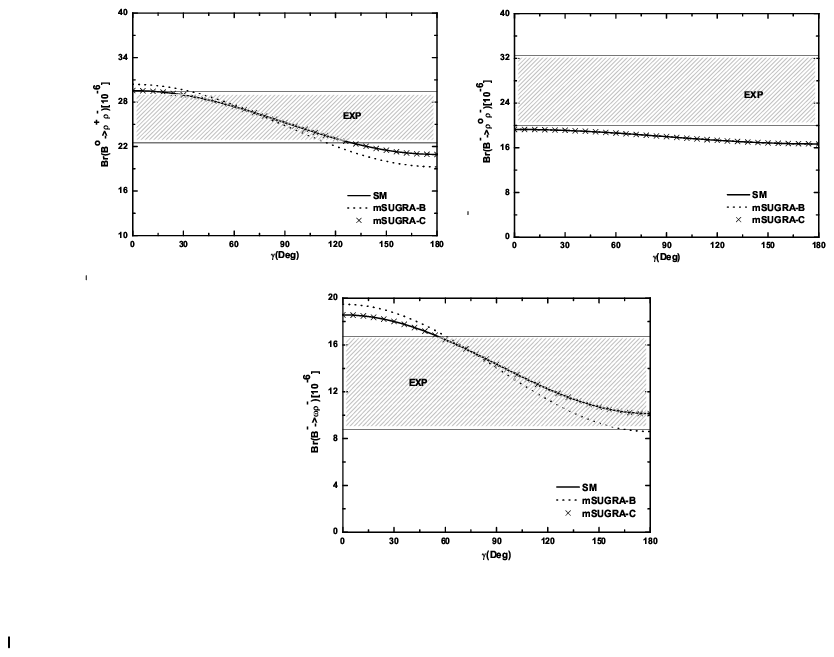}}}
\vspace{-1.5cm}
\caption{ The same as Fig.\ref{fig:pirho} but for $B^0 \to \rho^+\rho^-$ and
$B^-\to \rho^-(\rho^0, \omega)$ decays.}
\label{rho2}
\end{figure}

Fig.~\ref{rho2} shows the $\gamma$ dependence of the branching
ratios for the three decays in the SM and the mSUGRA model. From
this figure and the numerical results as given in
Eqs.(\ref{eq:rhorho1}-\ref{eq:rhow}), we can easily find that
the SUSY contributions to these tree-dominated decays are very small.
The largest error here still comes from the uncertainty of the relevant
form factors, and the theoretical predictions in both the SM and mSUGRA model are
consistent with the data within one standard deviation.

\subsubsection{ $B^{-}\to K^{*-}\rho^{0}$, $K^{*0}\rho^{-}$ and $B\to K^{*}\phi$}

\begin{figure}[tbp]
\vspace{-1cm}
\centerline{\mbox{\epsfxsize=20cm\epsffile{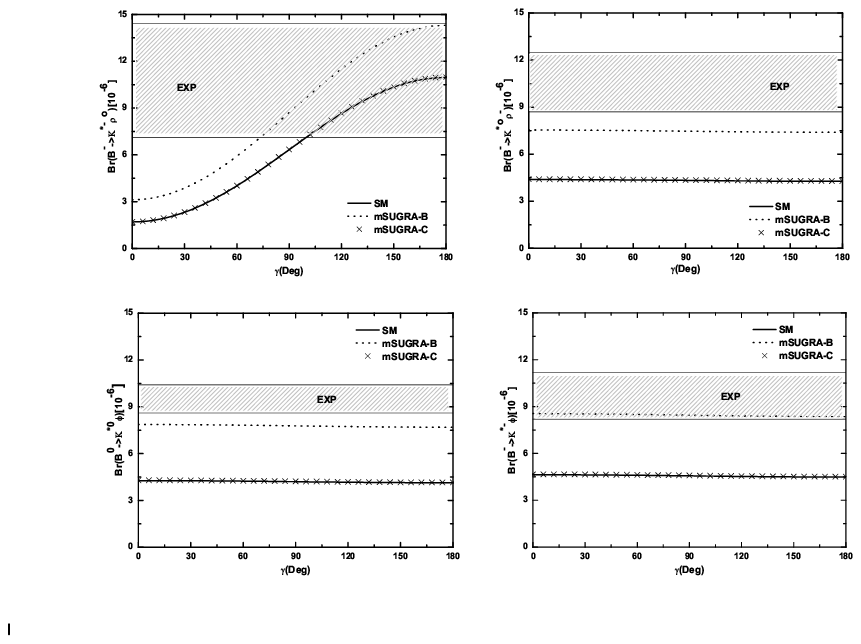}}}
\vspace{-1.5cm}
\caption{ The same as Fig.\ref{fig:pirho} but for $B^- \to K^{*-} \rho^0$,
$K^{*0}\rho^-$ and $B\to K^* \phi$ decays.}
\label{ksrho}
\end{figure}

Since these four decays are penguin dominated decay modes, the
SUSY contributions may be significant. The world averages,
the theoretical predictions (all in units of $10^{-6}$) in the SM and the
mSUGRA model are
\begin{eqnarray}
Br(B^ -  \to  K^{* - } \rho ^0) &=& \left \{\begin{array}{ll}
10.6 ^{+3.8}_{-3.5} , & {\rm  Data}, \\
3.9_{ - 0.5}^{ + 0.6} (\mu )\pm0.4 (A^{B \to \rho } )_{ - 0.4}^{ +
0.5} (A^{B \to K^* } )_{ - 1.2}^{ + 1.5} (\gamma ), & {\rm SM}, \\
{5.7_{ - 0.7}^{ + 0.9} (\mu )\pm0.6 (A^{B \to \rho } )\pm0.6
  (A^{B \to K^* } )_{ - 1.4}^{ + 1.8} (\gamma )},& {\rm Case-B}, \\
3.9_{ - 0.5}^{ + 0.6} (\mu )\pm0.4 (A^{B \to \rho } )_{ - 0.4}^{ +
0.5} (A^{B \to K^* } )_{ - 1.2}^{ + 1.5} (\gamma ),
   & {\rm Case-C}, \\
\end{array} \right. \label{eq:kprho1}\\
Br(B^ - \to  K^{*0} \rho ^ -)&=& \left \{\begin{array}{ll}
10.6 \pm 1.9 , & {\rm  Data}, \\
4.4_{ -0.9}^{ + 1.0} (\mu )_{ - 0.6}^{ + 0.7}
 (A^{B \to \rho } )_{ - 0.02}^{ + 0.01} (\gamma ),  & {\rm SM}, \\
 7.5_{ - 1.4}^{ + 1.6} (\mu )\pm1.1 (A^{B \to \rho } ) \pm 0.02(\gamma ),
 & {\rm Case-B}, \\
4.4_{ - 0.9}^{ + 1.0} (\mu )_{ - 0.6}^{ + 0.7}
 (A^{B \to \rho } )_{ - 0.02}^{ + 0.01} (\gamma ),
 & {\rm Case-C}, \\
\end{array} \right. \label{eq:kprho2}
\eeq
\beq
Br(\bar{B}^0 \to  \bar{K}^{*0}\phi)&=& \left \{\begin{array}{ll}
9.5 \pm 0.9 , & {\rm  Data}, \\
 4.2 ^{+1.4}_{-1.1}  (\mu) \;^{+1.7}_{-1.5} (A^{B\to K^*})\;
\pm0.02 (\gamma), & {\rm SM}, \\
7.8 ^{+2.1}_{-1.7}  (\mu) \; ^{+3.1}_{-2.6} (A^{B\to K^*})\;
 \;^{+0.02}_{-0.03}(\gamma),& {\rm Case-B}, \\
4.3 ^{+1.4}_{-1.1}  (\mu) \;^{+1.8}_{-1.5} (A^{B\to K^*})\;
\pm0.02 (\gamma),
& {\rm Case-C}, \\
\end{array} \right. \label{eq:kpo1}\\
Br(B^- \to  K^{*-}\phi)&=& \left \{\begin{array}{ll}
9.7 \pm 1.5 , & {\rm  Data}, \\
 4.6 ^{+1.5}_{-1.2}  (\mu) \;^{+1.9}_{-1.6} (A^{B\to K^*})\;
\pm0.02 (\gamma), & {\rm SM}, \\
 8.5 ^{+2.3}_{-1.8}  (\mu) \; ^{+3.4}_{-2.8} (A^{B\to K^*})\;
 \;^{+0.02}_{-0.03}(\gamma),& {\rm Case-B}, \\
4.6 ^{+1.5}_{-1.2}  (\mu) \;^{+1.9}_{-1.6} (A^{B\to K^*})\;
\pm0.02 (\gamma),
 & {\rm Case-C}. \\
\end{array} \right. \label{eq:kpo2}
\end{eqnarray}
Here the major errors are induced by the uncertainties of the input parameters
$\mu$, $A^{B \to \rho}$, $A^{B\to K^*}$ and the angle $\gamma$.

From the numerical results as given in
Eqs.(\ref{eq:kprho1}-\ref{eq:kpo2}) and Fig.~\ref{ksrho}, one can
see that
\begin{itemize}

\item
For these four channels, the central values of
the SM predictions are less than half of the measured values.
The SUSY contributions are negligibly small for the case C, but
can provide a $70\%$ enhancements for case C.

\item
The dominant errors still come from the uncertainty of
the form factors and the renormalization scale $\mu$.
Except for the decay $B^{-}\to K^{*-}\rho^{0}$,
other three decays are almost independent of the angle $\gamma$.
The reason is that the latter three decays are pure
penguin processes and have no terms proportional to $V_{td}$ or
$V_{ub}$.

\end{itemize}

\section{summary}

In this paper, we calculated the new physics contributions to the
CP-averaged branching ratios of the thirty nine $B\to PV$ and nineteen 
$B\to VV$ decay channels 
in the mSUGRA model by employing the QCD factorization approach.

In Sec.~\ref{sec:msugra}, a brief review about the mSUGRA model
and the SUSY corrections to the Wilson coefficients was given. In
Sec.~\ref{sec:qcdf}, we presented a short discussion about the QCD
factorization approach which is commonly used to evaluate the
hadronic matrix elements  $<M_1 M_2|O_i|B>$. In Sec.~\ref{sec:pv}
and Sec.~\ref{sec:vv},  we calculated the branching ratios of $B
\to PV$  and $B \to VV$ decays in the SM and the mSUGRA model, and
made phenomenological analysis for those well-measured decay
modes.

In the mSUGRA model, the Wilson coefficient $C_{7\gamma}(m_b)$ can
be either SM-like or sign-flipped comparing with that in the SM.
The data of $B \to X_s \gamma$ can provide a strong constraint on
the size of $C_{7\gamma}(m_b)$, but not its sign. The latest measurements
of $B \to X_s l^+ l^-$ decay prefer a negative ( SM-like) $C_{7\gamma}(m_b)$.
In this paper we choose three typical sets of the mSUGRA input parameters
$(m_0,m_{\frac{1}{2}},A_0,\tan{\beta},Sign(\mu))$ in which $C_{7\gamma}(m_b)$ 
can be either SM-like ( the case A and C) or
have a flipped-sign (the case B).

As expected, the SUSY contributions to all channels considered here
are very small for both the SM-like case A and C.
For case B, however, the SUSY contributions can be significant for those
penguin-dominated decays, and for this case we found that
\begin{itemize}

\item
For those tree-dominated decays or the decay channels having
the penguin contribution only coming from $\alpha_3,
\alpha_3^{ew}$ or having only weak annihilation contribution, the
SUSY contributions in case B  are also very small.

\item
For those QCD penguin-dominated decay modes, the SUSY contributions
can interfere with the corresponding SM parts constructively  or destructively,
and consequently can provide an enhancement about $30\% \sim
260\%$ to the branching ratios of the decays $B \to \pi K^*, K \phi, K^* \phi,
K^*\rho$ etc,
or a reduction about $30\% \sim 80\%$ to the branching ratios of the decays
$B\to K \rho, K\omega$ etc.

\item
For  $B\to K^*(\pi, \rho)$ and $B\to (K, K^*) \phi$ decays, the central 
values of the SM predictions for branching ratios are generally less than half
of the measured values. But the mSUGRA predictions can become consistent with the data
within one standard deviation due to the inclusion of
the significant SUSY enhancements.

\item 
For $B\to K \rho$ and $B\to K\omega$ decays, the SUSY
contributions interfere with their SM parts destructively, more
investigations are needed to cover the gap between the theoretical
predictions and the data.

\item 
For most decay channels, the error induced by the uncertainty of the annihilation 
contributions is dominant and large in size.
The SM predictions, consequently, have  large theoretical errors when the QCD 
factorization approach is employed.
The large SUSY contributions in the case B , unfortunately, may be masked 
by the large theoretical errors dominated by the uncertainty from
our ignorance of calculating the annihilation contributions
in the QCD factorization approach. Hence we expect the
theory will go further to make the annihilation contributions
calculated reliably and only at that time can one distinguish the
new physics contributions from the theoretical errors.

\end{itemize}

In short, only in case B, can the SUSY contributions be significant to the 
branching ratios of the studied B meson decays. 
If the sign of $C_{7\gamma}(m_b)$ can be strictly determined to be SM-like finally,
as claimed firstly in Ref.~\cite{pum05}, the parameter space of
the mSUGRA model will be strongly restricted, and the branching ratios of
the two-body charmless B decays can hardly be affected.

\begin{acknowledgments}

We are very grateful to Cai-dian L\"u and Li-bo Guo for helpful
discussions. This work is partially supported  by the National
Natural Science Foundation of China under Grant No.10275035 and 10575052, 
and by the Research Foundation of Nanjing Normal University under
Grant No.~214080A916.

\end{acknowledgments}


\begin{appendix}

\section{Input parameters}\label{sec:ip}

In this appendix we present the relevant input parameters being used in
our numerical calculations.

\begin{itemize}

\item
Decay constants. The decay constants are defined by following
current matrix elements\cite{bsw87}:
\begin{equation}
\left\langle {P(q)|\overline q  \gamma _\mu  \gamma _5 q |0}
\right\rangle  =  - if_p q_\mu\,, \qquad \left\langle
{V(q,\epsilon)|\overline q  \gamma _\mu q |0} \right\rangle =
 f_V m_V \epsilon^{*\mu} .
\end{equation}
And the scale-dependent transverse decay constant in
Eq.(\ref{eq:rchiV}) is defined as\cite{bn03b}
\begin{equation}
   \langle V(q,\varepsilon^*)|\bar{q}\sigma_{\mu\nu} q|0\rangle
   = f_V^\perp (q_\mu\varepsilon^*_\nu-q_\nu\varepsilon^*_\mu) \, ,
\end{equation}
where $\varepsilon^*_\mu$ denotes the polarization vector of the
vector meson V. These decay constants are nonperturbative
parameters. And they can be extracted from the experimental data or
estimated with well-founded theories, such as QCD sum rules, etc.
The decays constants we used here, in units of $MeV$,  are
\begin{center}
\begin{tabular}{ccc|cccc|cccc} \hline\hline
$f_\pi$ & $f_K$ & $f_B$& $f_{\rho}$ & $f_{K^*}$& $f_{\omega}$ &
$f_{\phi}$
& $f_{\rho}^\perp$ & $f_{K^*}^\perp$& $f_{\omega}^\perp$ & $f_{\phi}^\perp$ \\
\hline
131 & 160 & 200&209&218&187&221&150&175&150&175  \\
\hline
\end{tabular}
\end{center}

As to the decay constants related to $\eta$ and $\eta^{'}$, we
shall take the convention in Ref.~\cite{agc98}:
\begin{equation}
\langle {0|\overline q  \gamma _\mu  \gamma _5 q |\eta^{(')}(p)}
\rangle  =  - if_{\eta^{(')}}^{q} p_\mu
\end{equation}
the quantities $f^{u}_{\eta^{(')}}$ and $f^{s}_{\eta^{(')}}$ in
the two-angle mixing formalism are
\begin{eqnarray}
 f_{\eta^{'}}^u  = \frac{{f_8 }}{{\sqrt 6 }}\sin \theta _8  + \frac{{f_0 }}{{\sqrt 3 }}\cos \theta _0 \,, \qquad
 f_{\eta^{'}}^s  =  - 2\frac{{f_8 }}{{\sqrt 6 }}\sin \theta _8  + \frac{{f_0 }}{{\sqrt 3 }}\cos \theta _0  \\
 f_\eta ^u  = \frac{{f_8 }}{{\sqrt 6 }}\cos \theta _8  - \frac{{f_0 }}{{\sqrt 3 }}\sin \theta _0 \,, \qquad
 f_\eta ^s  =  - 2\frac{{f_8 }}{{\sqrt 6 }}\cos \theta _8  - \frac{{f_0 }}{{\sqrt 3 }}\sin \theta _0
\end{eqnarray}
with $f_8=1.28f_{\pi}$, $f_0=1.2f_{\pi}$, $\theta_0=-9.1^o$ and
$\theta_8=-22.2^o$. In order to incorporate the charm-loop diagram
contribution into $\eta^{(')}$ involved decays, two new
decay constants should be given, namely $f^c_{\eta}$ and
$f^c_{\eta^{'}}$. Following Ref.~\cite{bn03a}, we take  values
$f^c_{\eta}\simeq-1MeV$ and $f^c_{\eta^{'}}\simeq-3MeV$.

\item Form factors. The form factors are in nature nonperturbative
quantities, extracted usually from the experimental measurements.
In Ref.~\cite{bsw87}, the form factors were defined by current
matrix elements
 \begin{eqnarray}
 {\langle}P(q){\vert}\bar{q}{\gamma}^{\mu}
       (1-{\gamma}_{5})q{\vert}B{\rangle}
    &=& \Big[ p_{B}^{\mu} + q^{\mu}
     - \frac{m_{B}^{2}-m_{P}^{2}}{k^{2}} k^{\mu}
       \Big] F_{1}(k^{2})\nonumber \\
     &+& \frac{m_{B}^{2}-m_{P}^{2}}{k^{2}} k^{\mu} F_{0}(k^{2}), \\
 {\langle}V(q,{\epsilon}){\vert}\bar{q}{\gamma}_{\mu}
       (1-{\gamma}_{5})q{\vert}B{\rangle}
    &=& {\epsilon}_{{\mu}{\nu}{\alpha}{\beta}}{\epsilon}^{{\ast}{\nu}}
       p_{B}^{\alpha} q^{\beta} \frac{2V(k^{2})}{m_{B}+m_{V}}
     + i \frac{2m_{V}({\epsilon}^{\ast}{\cdot}k)}{k^{2}}
       k_{\mu} A_{0}(k^{2}) \nonumber \\
    &+& i {\epsilon}^{\ast}_{\mu} ( m_{b}+m_{V} ) A_{1}(k^{2})
     - i \frac{{\epsilon}^{\ast}{\cdot}k}{m_{b}+m_{V}}
          ( p_{B} + q )_{\mu} A_{2}(k^{2}) \nonumber \\
    &-& i \frac{2m_{V}({\epsilon}^{\ast}{\cdot}k)}{k^{2}}
       k_{\mu} A_{3}(k^{2}),
 \end{eqnarray}
where $k=p_{B}-q$.  We neglect corrections to
the decay amplitudes quadratic in the light meson masses, so that
all form factors are evaluated at $k^2=0$. At the poles $k^{2}=0$,
we have
\beq
F_{0}(0)&=&F_{1}(0), \quad  A_{0}(0)=A_{3}(0), \\
2 m_{V} A_{3}(0) &=& (m_{B}+m_{V}) A_{1}(0) - (m_{B}-m_{V}) A_{2}(0).
 \label{eq:conditions}
 \eeq

In our calculations, we use the form factors as given in
Refs.~\cite{bn03b,pr05}. They are
\begin{center}
\begin{tabular}{c|ccccc} \hline\hline
\ \ &$B\to\pi$&$B\to K$&$B\to\rho$&$B\to K^*$&$B\to\omega$\\
\hline
$F_0$&$0.28\pm0.05$&$0.34\pm0.05$&$-$ & $-$& $-$\\
$A_0$&$-$ & $-$ &$0.37\pm0.06$&$0.45\pm0.07$&$0.33\pm0.05$\\
$A_1$&$-$ & $-$  &$0.242\pm0.02$ &$0.292\pm0.06$&$0.219\pm0.02$\\
$A_2$&$-$ & $-$  &$0.221\pm0.02$ &$0.259\pm0.06$&$0.198\pm0.02$\\
V    &$-$ & $-$  &$0.323\pm0.03$ &$0.411\pm0.08$&$0.293\pm0.03$\\
\hline\hline
\end{tabular}
\end{center}

For the form factors of $B\to \eta$ and $B\to \eta^{'}$
transitions, we use \cite{agc98}
\beq
F_{0,1}^{B \to \eta }  = F_{0,1}^{B \to \pi } (\frac{\cos \theta
_8}{\sqrt 6 } - \frac{\sin \theta _0 }{\sqrt 3 }), \qquad
F_{0,1}^{B \to \eta ^{'} }  = F_{0,1}^{B \to \pi } (\frac{\sin
\theta _8 }{\sqrt 6 } + \frac{\cos \theta _0 }{\sqrt 3 }).
\eeq

\item
The parameters $\chi_{H,A}$. When calculating the
contribution from hard spectator scattering and annihilation
diagrams, the end point divergence will appear. To treat the
divergence, one can parameterize them by $\chi_{H}$ and $\chi_{A}$
respectively \cite{bbns01}.

In numerical calculations, we set $\rho_H=\rho_A=0$ as the default
input values \cite{bbns01}. And we also scan over $\rho_A\in[0,1]$ and
$\phi_A\in[-180^\circ, 180^\circ]$ to estimate the theoretical
error induced by the uncertainty of the annihilation contributions.

\item
In this paper, we use the same CKM angles, quark masses,
meson masses and the B meson lifetimes as used in Ref.~\cite{zw04}.

\end{itemize}

\end{appendix}


\end{document}